\documentclass[11pt, a4paper]{article}

\usepackage{color}
\usepackage{graphicx}
\usepackage{amssymb}
\usepackage{amsmath}
\usepackage[english]{babel}
\usepackage[T1]{fontenc}
\usepackage[noadjust]{cite}
\usepackage{hyperref} 

\numberwithin{equation}{section}

\usepackage{booktabs}
\usepackage{mdframed}
\usepackage{xcolor}

\usepackage{bbold}
\usepackage[utf8]{inputenc}
\usepackage{authblk}

\newcommand\nn{\nonumber}
\newcommand\eeq{\end{equation}}
\newcommand\beq{\begin{equation}}
\newcommand\eea{\end{eqnarray}}
\newcommand\bea{\begin{eqnarray}}

\begin{document}

\linespread{1.1}

\title{{\bf Gauge coupling unification and doublet-triplet splitting
via GUT dynamical breaking}}

\author[1,2]{ {\large\sc Isabella Masina} \thanks{masina@fe.infn.it}}
\author[3]{ {\large\sc Mariano Quir\'os} \thanks{quiros@ifae.es}}

\affil[1]{\small\it Dept.\,of Physics and Earth Science, Ferrara University, Via Saragat 1, 44122 Ferrara, Italy }
\affil[2]{\small\it Istituto Nazionale di Fisica Nucleare (INFN), Sez.\,di Ferrara, Via Saragat 1, 44122 Ferrara, Italy }
\affil[3]{\small\it Institut de F\'{\i}sica d'Altes Energies (IFAE) and The Barcelona Institute of  Science and Technology (BIST), Campus UAB, 08193 Bellaterra (Barcelona) Spain}

\date{}

\maketitle
\begin{abstract}
An interesting framework to achieve gauge coupling unification consists in adding to the Standard Model Lagrangian non-renormalizable operators of $d \geq 5$, which affect the kinetic term of gauge fields. We first review the phenomenology related to this framework in the context of $SU(5)$, identifying which are the most interesting representations for the sake of achieving coupling unification. 
Secondly, we point out that in the case of a dynamical breaking pattern, it is possible to relate gauge coupling unification with the doublet-triplet splitting problem. 
We show that condensates of fermions in the $5$ representation do not lead to viable models because of proton decay constraints. 
At difference, we point out that successful models can be obtained by considering condensates of fermions in the $10$, as well as in the $24$ representations. 
\end{abstract}

\linespread{1.3}

\vskip 1.cm
\newpage

\section{Introduction}

The beta functions of the Standard Model (SM) of particle interactions are such that gauge coupling unification (GCU) is slightly, but unavoidably, missed. 
In a previous related work, Ref.\,\cite{Masina:2025klj}, we discussed a simple general parameterization for the new physics corrections leading to full unification at some scale $M_X$. 
We showed that for any new physics model such that the corrections to the non-Abelian couplings are equal (or nearly so), $M_X$ is equal (or close to) the partial
unification scale of the SM non-Abelian running gauge couplings, $\mu_{32}^{\rm SM}  \approx 2.8 \times 10^{16}$ GeV. 
The latter scales can be disentangled only if the corrections to the non-Abelian couplings are significantly different. 

We already explored in~\cite{Masina:2025klj} how the parameterization works for some relevant models without a desert up to the scale $M_X$,
as low energy supersymmetry (SUSY), and many others. 
For models with a desert, GCU relies on corrections at $M_X$ which are instead of ultraviolet (UV) origin. 
Examples of UV origin corrections include: 
string inspired corrections~\cite{Dienes:1996du, Cho:1997gm}, also explored in\,\cite{Masina:2025klj}, and
effective corrections induced by an additional non-renormalizable kinetic term in the Lagrangian~\cite{Hill:1983xh, Shafi:1983gz, Panagiotakopoulos:1984wf, Hall:1992kq}.
In the present work we complete the analysis performed in \cite{Masina:2025klj}, by focussing on the latter case, whose natural embedding is within a Grand Unified Theory (GUT)~\cite{Georgi:1974sy}.

In a GUT framework, gauge couplings near the unification scale can receive corrections, that are mainly of two kinds: 
\textit{i)} corrections related to the mass spectrum of GUT particles, and \textit{ii)} the previously mentioned corrections of UV origin. 
\begin{itemize}
\item
As for \textit{i)}, gauge couplings can receive threshold corrections from GUT particles if the latter have masses below the GCU scale, as their presence modifies the beta functions. 
These corrections are in general subdominant and cannot alone account for GCU. For an example of this approach see e.g.~\cite{DiLuzio:2013dda, Haba:2024lox}. 
\item
As for \textit{ii)}, gauge couplings can be modified as an effect of non-renormalizable $d \geq 5$ operators which affect the kinetic term of gauge fields \cite{Hill:1983xh, Shafi:1983gz, Panagiotakopoulos:1984wf, Hall:1992kq}. They might arise by embedding the GUT in a larger one, or because of gravitational effects. 
In any case, the suppression scale of the non-renormalizable operators has to be associated with the GUT's cutoff scale, $\Lambda$ (for instance, identified with the Planck scale, $M_{P}$, in \cite{Hill:1983xh}, or with the compactification scale in \cite{Shafi:1983gz}).  These UV origin corrections, which alone might account for GCU, 
are expected to be small and one would in general expect them to be treated perturbatively. 
\end{itemize}

In this work, we firstly reconsider the scenario 2) in the framework of $SU(5)$ and in relation with the parameterization proposed in\,\cite{Masina:2025klj}. 
Secondly, we discuss its relation with another relevant issue of GUT, that is the doublet-triplet splitting (DTS) problem.
In particular, we propose that, in the case that the GUT breaking is dynamical, GCU and DTS are related and can be both addressed successfully.

Let us now summarize the status of the art about GCU from non-renormalizable $d \geq 5$ operators.
Since 1984, it has been pointed out, Refs.~\cite{Hill:1983xh, Shafi:1983gz, Panagiotakopoulos:1984wf, Hall:1992kq}, 
that $d=5$ operators involving Higgs fields responsible for breaking $SU(5)$ into the SM do affect GCU, and that GCU can be achieved in this way; 
the related problem of proton decay was also considered. 
The pioneering studies focused on the $24$ representation, including Ref.\,\cite{Calmet:2008df}, where the role of an effective Planck scale was emphasized; 
in Ref.\,\cite{Bhatt:2008qb} also $d=6$ operators where studied; Refs.\,\cite{Chakrabortty:2008zk, Calmet:2009hp} extended the calculation to the $75$ and $200$ representations.
These previous numerical analyses have however not been conclusive about the emerging GUT pattern. 

In the present work, with a bottom-up attitude, we begin by reconsidering the latter issue, and find that an analytical approach helps in grasping which are the most interesting directions that are supported by phenomenological data, and which are the most relevant ones for model building. 
In particular, we study and highlight the differences between the scenarios in which the breaking of $SU(5)$ is realized by means of the $24$ representation, or rather the $75$ and $200$ representations.
For instance, as a phenomenologically remarkable scenario, we discuss the one in which both the $24$ and $75$ are present and, as an effect of the relative magnitude of their vacuum expectation values (VEVs), GCU is achieved at the same scale where the SM non-Abelian gauge couplings unify\,\footnote{Since the latter scale is close to the typical unification scale in the framework of low energy supersymmetry, this scenario can be dubbed "mirage SUSY" (mS), as proposed in Ref.~\cite{Masina:2025klj}.}. 
We provide a short  discussion on the issue of proton decay for the scenarios that we find more interesting.

We then investigate the possible relation between GCU and DTS. In the case that the $SU(5)$ breaking is realized via a Higgs mechanism, there is actually no relation. 
On the contrary, we point out that in the case of a dynamical breaking of $SU(5)$, it is possible to relate the two issues, and even simultaneously address them successfully. 
We find that the most interesting possibilities are those where the effective representations involved in the $d=5$ operator arise from a fermion condensate with fermions sitting in the $10$ or $24$ representations; the case of the $5$ representation being ruled out by proton decay constraints. 

The paper is organized as follows.
In Sec.\,\ref{sec-GCU} we review and update the analysis of GCU from $d=5$ operators in an $SU(5)$ context. 
Sec.\,\ref{sec-BU} is devoted to the analysis of some relevant scenarios, also in connection with proton decay.
In Sec.\,\ref{sec-DTS} we discuss the relation between GCU and DTS, which follows as a consequence of dynamical breaking. 
The phenomenology of the latter scenario is studied in detail in Sec.\,\ref{sec-PD}.
Conclusions are drawn in Sec.\,\ref{sec-conc}. 
App.\,\ref{appgroup} contains technical material about group theory, useful to clarify the calculations presented in the text.
App.\,\ref{App-app} deals with a useful analytical approximation to study the GCU corrections induced by $d=5$ kinetic operators.  
Finally, in App.\,\ref{apptoy} a toy UV model for non-renormalizable operators is proposed.

\section{GCU from $d=5$ operators in $SU(5)$ }
\label{sec-GCU}

We consider an $SU(5)$ GUT~\cite{Georgi:1974sy} with fermion matter fields in the $10_F$ and $\bar 5_F$ representations. 
The gauge fields are in the adjoint representation, $24_G$.
 As for scalar fields, the SM Higgs doublet is assumed to be in the $5_H$, and its conjugate in the $\bar 5_H$. 
 For a review of the notation, see App.~\ref{appgroup}.

The kinetic term of the gauge bosons is proportional to ${\rm Tr}(G_{\mu\nu} G^{\mu\nu})$, where $G_{\mu\nu} \equiv G_{\mu\nu}^A \lambda_A$ is the field strength tensor, $\lambda_A$ are the generators of $SU(5)$ in the adjoint representation, and the sum over $A$ is understood. 
From the group theoretical point of view, $G_{\mu\nu} G^{\mu\nu}$ is the symmetric part of the product of two adjoints, that is $(24\times 24)_s =1_s+24_s+75_s+200_s$. 
Taking the trace amounts to considering the singlet $1_s$.

As pioneered in \cite{Hill:1983xh, Shafi:1983gz, Panagiotakopoulos:1984wf}, a correction to the gauge couplings is induced 
by adding to the standard kinetic term the non-renormalizable $d=5$ operator 
\beq
{\mathcal L}_G + \delta {\mathcal L}_G 
= -\frac{1}{4} \, {\rm Tr}(G_{\mu\nu} G^{\mu\nu}) -\frac{1}{4} \, \sum_r \frac{c_r}{\Lambda} {\rm Tr}(G_{\mu\nu} G^{\mu\nu} H_r) \,\, ,
\label{eq-opgr}
\eeq
where the Higgs $H_r$ is a scalar field in the $r$-dimensional irreducible representation (irrep) of $SU(5)$, for which the allowed values are: $r=1, 24, 75, 200$. 
Notice also that $H_r$ might be a propagating field, 
as well as a non-propagating one in an effective theory\,\footnote{For instance, effective representations are obtained from the product $10\times \overline{ 10} =1+24+75$.}; for the sake of the following analysis, the possibility of having propagating or effective representations makes no difference. 
Indeed, in any case $H_r$ will acquire a VEV, spontaneously or dynamically breaking the GUT (more on this later). 

The $d=5$ operator in (\ref{eq-opgr}) leads to the modifications in the kinetic terms of the SM gauge fields 
\beq
- \frac{1}{4} (1+\epsilon_1 ) ( F_{\mu\nu} F^{\mu\nu}   )_{U(1)} - \frac{1}{2} (1+\epsilon_2 ) ( F_{\mu\nu} F^{\mu\nu}   )_{SU(2)}
- \frac{1}{2} (1+\epsilon_3 ) ( F_{\mu\nu} F^{\mu\nu}   )_{SU(3)} \,\, .
\eeq
Upon gauge fields redefinitions leading to canonical kinetic terms, the relations expressing GCU at the scale $\mu=M_X$ become
\beq
\alpha_G= (1+\epsilon_1 ) \,\alpha_1(M_X)=(1+\epsilon_2 )\, \alpha_2(M_X)=(1+\epsilon_3 ) \,\alpha_3(M_X)\,,
\label{eq-unif-gr}
\eeq
where the SM running gauge couplings $\alpha_s(\mu)$, with $s=1,2,3$, are obtained via RGE running from those measured at low energy.
For the irreps in Eq.~(\ref{eq-opgr}), the corrections in Eq.~(\ref{eq-unif-gr}) are given by the sum of four possible contributions \cite{Calmet:2009hp, Chakrabortty:2008zk} 
\beq
\epsilon_s = \sum _{r}\epsilon_s^{(r)}  \,\, \,, 
\label{eq-epsilons}
\eeq
where $s=1,2,3$ and the following relations (see for instance Table 2 of Ref.~\cite{Calmet:2009hp}),
\beq
\epsilon_1 = \alpha +\beta +\gamma  +\delta \,\, ,\,\,  \epsilon_2=  \alpha + 3 \beta -\frac{3}{5}\gamma +\frac{1}{5}\delta \,\,  , \,\,  \epsilon_3= \alpha -2 \beta -\frac{1}{5}\gamma +\frac{1}{10}\delta \,\,.
\eeq
hold for~\footnote{We find it useful this notation, as the increasing alphabetic ordering of the Greek letters corresponds to increasing the dimension of the irreps.}
\bea
&\,&\alpha \equiv \epsilon_1^{(1)} =\epsilon_2^{(1)} =\epsilon_3^{(1)}  \nn \\
&\,&\beta  \equiv \epsilon_1^{(24)} =\frac{1}{3} \epsilon_2^{(24)} = - \frac{1}{2} \epsilon_3^{(24)} \nn \\
&\,&\gamma  \equiv \epsilon_1^{(75)} = -\frac{5}{3} \epsilon_2^{(75)} = - 5 \,\epsilon_3^{(75)} \nn \\
&\,&\delta  \equiv \epsilon_1^{(200)} = 5 \,\epsilon_2^{(200)} = 10\, \epsilon_3^{(200)}  \, \, .
\label{eq-abcd}
\eea

As for the absolute size of the corrections, we have to define quantities related to $SU(5)$ breaking.
For instance, for the VEVs of $H_1$ and $H_{24}$, we can exploit a $5\times 5$ matrix representation, 
\beq
\langle H_1 \rangle= v_1\, {\mathbb{1}}_5 \,   \,\,\, , \,\,\, \langle H_{24} \rangle =v_{24}\, {\rm diag}(\mathbb{1}_3,-3/2 \cdot \mathbb{1}_2)  \,\, ,
\eeq
so that the breaking to the SM is achieved provided $v_{24} \neq 0$. 
According to our notations, as discussed in App.\,\ref{appgroup}, the corrections to e.g.~$\epsilon_3$ can be cast in the form
\beq
\epsilon^{(r)}_3=  \frac{c_r v_r} {\Lambda} \,\, , 
\eeq
for any possible value of $r=1,24,75,200$.

\subsection{Our method}

Here we reconsider the phenomenological analysis about achieving GCU from a non-renormalizable kinetic term. Such an issue can be addressed and solved exactly, as well as in a semi-analytical way.

\begin{figure}[htb!]
\vskip .5cm 
 \begin{center}
\includegraphics[width=7 cm]{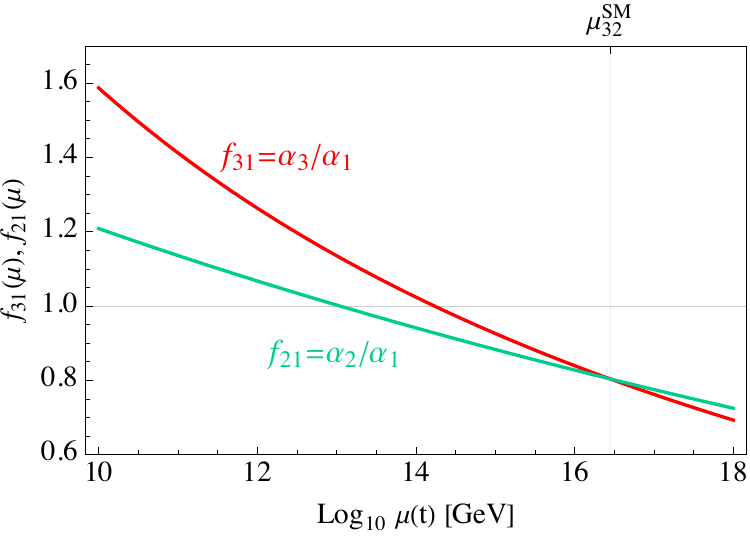}  
 \end{center}
\caption{\baselineskip=12 pt \small \it
The quantities $f_{21}(\mu) \equiv \frac{\alpha_2(\mu)}{ \alpha_1(\mu)} $ and $ f_{31}(\mu) \equiv \frac{\alpha_3(\mu)}{ \alpha_1(\mu)}$, from a calculation at NNLO.} 
\label{fig-f3121}
\vskip .5 cm
\end{figure}

Let us start by putting Eq.~(\ref{eq-unif-gr}) in the equivalent form
\beq
1= \frac{ 1+ \alpha + 3 \beta -\frac{3}{5}\gamma +\frac{1}{5}\delta }{1+\alpha +\beta +\gamma  +\delta} \frac{\alpha_2(M_X)}{ \alpha_1(M_X)} 
= \frac{1+ \alpha -2 \beta -\frac{1}{5}\gamma +\frac{1}{10}\delta}{1+\alpha +\beta +\gamma +\delta }  \frac{\alpha_3(M_X)}{ \alpha_1(M_X)}   \,\,,
\label{eq-unif-gr-2}
\eeq
which emphasizes the role of ratios of the SM running gauge couplings,  
\beq
f_{21}(\mu) \equiv \frac{\alpha_2(\mu)}{ \alpha_1(\mu)} \,\, , \,\,\,  f_{31}(\mu) \equiv \frac{\alpha_3(\mu)}{ \alpha_1(\mu)} \,\, .
\eeq  
The running ratios above can be determined with high accuracy by performing the calculation at NNLO (which requires to exploit the RGE and the matching conditions at least at 3-loops and 2-loops respectively), as shown in Fig.~\ref{fig-f3121}. 
Notice that, as already mentioned and discussed in detail in Ref.\,\cite{Masina:2025klj}, the non-Abelian gauge couplings unify at the scale $\mu_{32}^{\rm SM}\approx 2.8 \times 10^{16}$ GeV,
where they take the value $\alpha_{32}^{\rm SM} \approx 0.0217$.

The system in Eq.\,(\ref{eq-unif-gr-2}) can be solved exactly, as we will do in the following, 
or within an approximation, as discussed in App.\,\ref{App-app}. 
Let us define 
\beq
\tilde \beta=\beta/(1+\alpha) \,, \,\,    \tilde \gamma= \gamma/(1+\alpha)\,, \,\, \tilde \delta= \delta/(1+\alpha) \,.
\eeq 
The exact solution of the system in Eq. (\ref{eq-unif-gr-2}) is then given by
\bea
\tilde\beta&=&\frac{f_{21}f_{31}(20+\tilde\delta)+15 f_{31}(4+\tilde\delta)-40 f_{21}(2+\tilde\delta)}{90[f_{31}+f_{21}(2+f_{31})]}\nonumber\\
\tilde\gamma&=&\frac{f_{21}f_{31}(50+7\tilde\delta)-3 f_{31}(10+7\tilde\delta)-4 f_{21}(5+7\tilde\delta)}{18[f_{31}+f_{21}(2+f_{31})]} \,\,   .
\label{eq-tbtg-ex}
\eea
The quantities $\tilde \beta$ and $\tilde \gamma$ are shown in the left plot of Fig.~\ref{fig-tbtg}, for selected values of $\tilde \delta$. 
While $\tilde \gamma$ strongly depends on $\tilde \delta$, this is not the case for $\tilde \beta$, for which the dependence is mild.
Notice also that one cannot phenomenologically determine $\alpha$: 
its variation corresponds to an overall increase or decrease of $\alpha_G$, and not to a shift of the GCU scale $M_X$. 

In the remaining part of this section, we will first ignore $H_{200}$ (thus taking $\delta =0$), and explore the contributions from $H_{1,24,75}$.
Secondly, (for reasons that will be clarified later) we will instead ignore $H_{24}$ (thus taking $\beta =0$), and explore the contributions from $H_{1,75, 200}$.

\subsection{Analysis of the contributions from $1$, $24$ and $75$}
\label{subsec-75}

\begin{figure}[htb!]
\vskip .5cm 
 \begin{center}
 \includegraphics[width=7.6 cm]{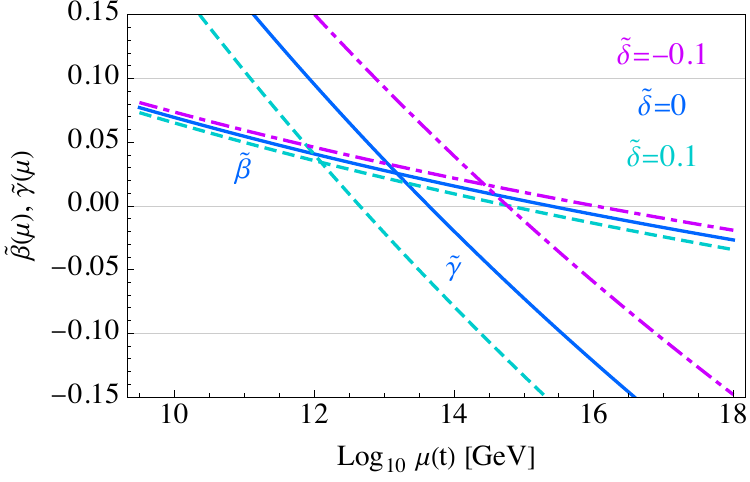}   \,\,  \includegraphics[width=7.8 cm]{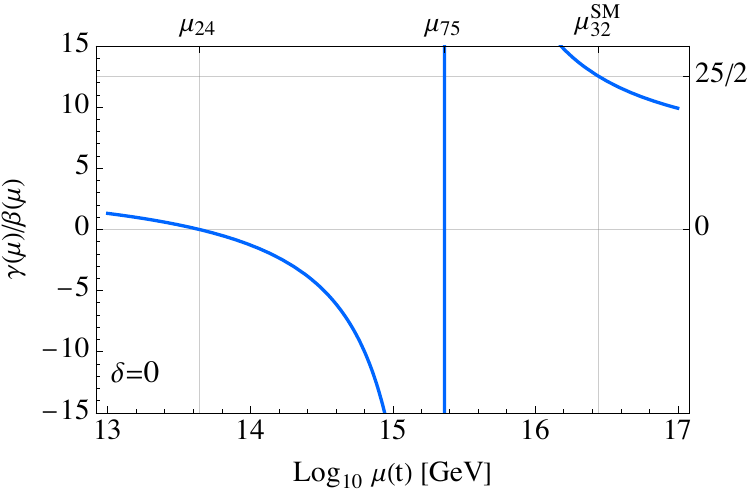}  
 \end{center}
\caption{\baselineskip=12 pt \small \it
Left: Solid lines are $\tilde \beta(\mu)$ and $\tilde \gamma(\mu)$ according to the exact solution, for $\tilde \delta=0, \pm 0.1$.
Right: Exact solution for the ratio $\gamma(\mu)/\beta(\mu)$ with $\delta =0$.} 
\label{fig-tbtg}
\vskip .5 cm
\end{figure}

If we only allow for the presence of the representations $1,24$ and $75$, 
the two unknowns $\tilde \beta$ and $\tilde \gamma$, can be univocally determined as a function of $\mu$, as shown by the solid lines in Fig.~\ref{fig-tbtg},
obtained using the exact solution (\ref{eq-tbtg-ex}) with $\tilde \delta=0$.  
First of all, we can check that for any scale $\mu$ in the range $10^{11.5}-10^{16.5}$ GeV there are solutions such that both $|\tilde \beta|$ and $|\tilde \gamma|$ are small (as assumed), say smaller than $0.15$. So, GCU can be reasonably be achieved for $\mu$ in such a range.

We now focus on two particular cases. One can see that it is possible to realize GCU exploiting solely the $24$: 
this happens for $\tilde \gamma=0$ (no 75) and $\tilde\beta\approx 0.02$, in which case $M_X =\mu_{24} \approx 10^{13.6}$ GeV. 
On the other hand, it is possible to realize GCU also solely with the $75$: this happens for $\tilde \beta=0$ (no 24) and $\tilde \gamma \approx -0.1$, in which case $M_X =\mu_{75}\approx 10^{15.4}$ GeV.

It is also interesting to inspect the ratio $\tilde \gamma (\mu)/\tilde \beta (\mu)=\gamma (\mu)/\beta (\mu)$, 
as shown in the right plot of Fig.~\ref{fig-tbtg}.  
Clearly, the ratio is vanishing at $\mu_{24}$ and diverges at $\mu_{75}$.
An interesting value for the ratio is $25/2$, which corresponds to $\mu=\mu_{32}^{\rm SM} \approx 2.8 \times 10^{16}$ GeV. 

\begin{figure}[htb!]
\vskip .5cm 
 \begin{center}
\includegraphics[width=7.4 cm]{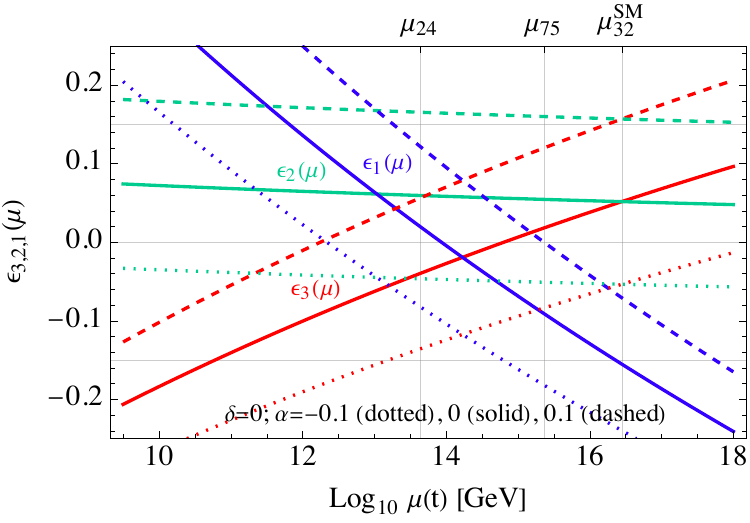} \,\,\,\,\, \includegraphics[width=7.4 cm]{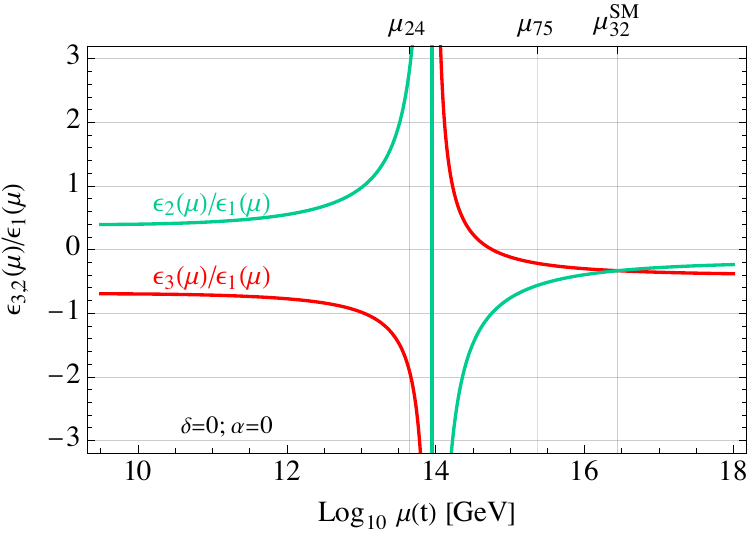} 
 \end{center}
\caption{\baselineskip=12 pt \small \it
Left: Dotted, solid, dashed lines are $\epsilon_s(\mu)$ for $\alpha=-0.1,0,0.1$ respectively. Right: ratios $\epsilon_{2,3}(\mu)/\epsilon_1(\mu)$, with $\alpha=0$.
} 
\label{fig-eps}
\vskip .5 cm
\end{figure}

The functions $\epsilon_i(\mu)$, obtained by substituting in Eq.\,(\ref{eq-epsilons}) the exact expressions for $\tilde \beta(\mu)$ and $\tilde \gamma(\mu)$, Eq.~(\ref{eq-tbtg-ex}),
are shown in the left plot of Fig.~\ref{fig-eps}, for $\alpha=-0.1$ (dotted), $0$ (solid), $0.1$ (dashed). 
Taking in particular $\alpha=0$, the right plot shows the ratios $\epsilon_{2,3}(\mu)/\epsilon_1(\mu)$.

Exploiting Fig.~\ref{fig-eps} and taking $\alpha=0$, we now discuss a few interesting cases for GCU
(the extension to $\alpha \neq 0$ will be discussed in the next section). 

\begin{figure}[htb!]
\vskip .5cm 
 \begin{center}
\includegraphics[width=7 cm]{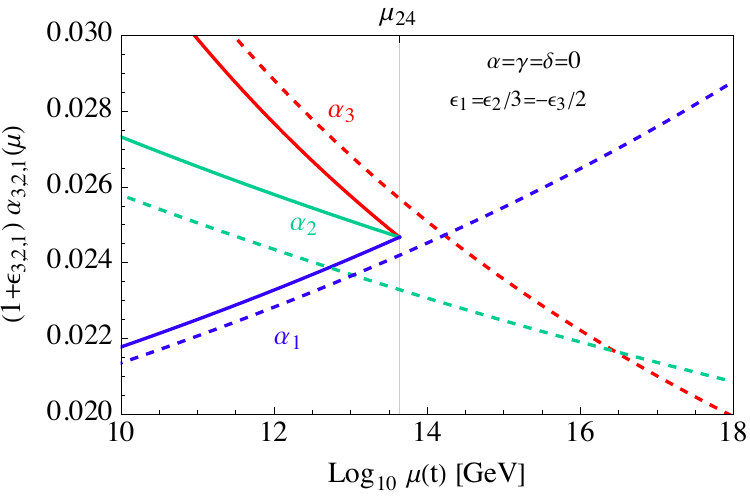}  \,\,\,\,\, \,\,\,\,\,\,\includegraphics[width=7 cm]{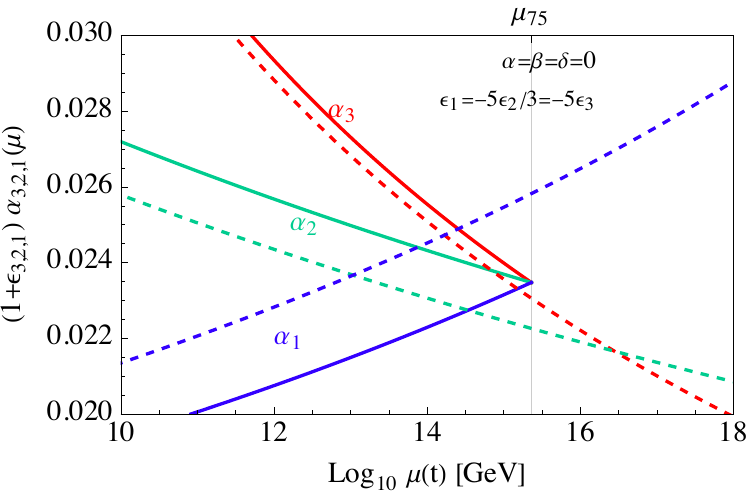} 
\,\, \includegraphics[width=7 cm]{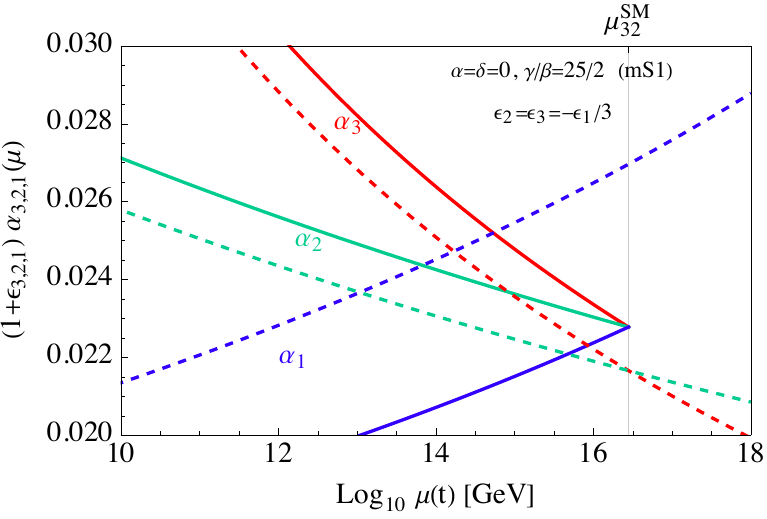}   \end{center}
\caption{\baselineskip=12 pt \small \it
Top Left: GCU happens at $\mu_{24}$  by taking $0.020=\epsilon_1(\mu_{24}) \approx \epsilon_2(\mu_{24})/3  \approx - \epsilon_3(\mu_{24})/2$.  
Top Right: GCU happens at $\mu_{75}$  by taking $-0.091\approx  \epsilon_1(\mu_{75}) \approx -5\epsilon_2(\mu_{75})/3  \approx - 5 \epsilon_3(\mu_{75})$. 
Bottom: GCU happens at $\mu_{mS}$ by taking $0.052\approx \epsilon_2(\mu_{mS})\approx \epsilon_3(\mu_{mS}) \approx -\epsilon_1(\mu_{mS})/3$.
Dashed lines are the gauge couplings in the SM. } 
\label{fig-unif}
\vskip .5 cm
\end{figure}

\begin{itemize}
\item $24$ only ($\alpha=\gamma=\delta=0$): $M_X=\mu_{24}\approx 10^{13.6}$ GeV. \\ 
By taking
\beq
0.020 \approx \beta =  \epsilon_1(\mu_{24}) \approx \frac{1}{3} \epsilon_2(\mu_{24})  \approx - \frac{1}{2} \epsilon_3(\mu_{24})   \,\, ,
\label{eq-mu24a0}
\eeq
GCU happens at $\mu_{24}$ with $\alpha_G \approx 0.0247$, as shown in the top left plot of Fig.~\ref{fig-unif}. 
The solid lines unifying at $\mu_{24}$ represent the combinations $(1+\epsilon_s) \alpha_s(\mu)$, 
while dashed lines are the SM running couplings, $\alpha_s(\mu)$.

\item $75$ only ($\alpha=\beta=\delta=0$): $M_X=\mu_{75}\approx 10^{15.4}$ GeV. \\ 
By taking 
\beq
-0.091\approx  \gamma= \epsilon_1(\mu_{75}) \approx -\frac{5}{3} \epsilon_2(\mu_{75})  \approx - 5 \epsilon_3(\mu_{75}) \, ,
\label{eq-mu75a0}
\eeq 
GCU happens at $\mu_{75}$ with $\alpha_G \approx 0.0235$, as shown in the top right plot of Fig.~\ref{fig-unif},

\item $mS_1$ ($\alpha=\delta=0$, $\gamma/\beta=25/2$): $M_X=\mu^{\rm SM}_{32}=\mu_{mS}=10^{16.4}$ GeV $\approx 2.8\times 10^{16}$ GeV. \\
By taking
\beq
0.052\approx \epsilon_2(\mu_{mS})\approx \epsilon_3(\mu_{mS}) \approx -\epsilon_1(\mu_{mS})/3 \,\,\, ,
\label{eq-muSa0}
\eeq 
GCU is achieved at the scale $\mu_{32}^{\rm SM}$ with $\alpha_G \approx 0.0228$, as shown in the bottom plot of Fig.~\ref{fig-unif}, 
Since $\mu_{32}^{\rm SM}$ is very close to the GCU scale with low energy supersymmetry, this scenario can be denoted as mirage SUSY ($mS_1$) \,\cite{Masina:2025klj};
hence we define $\mu_{mS} \equiv \mu_{32}^{\rm SM}$. 
We find it interesting that the values of the $\epsilon_s$ (obtained with just the 24 and 75 irreps) mimicking low energy SUSY correspond to the very simple ratios of Eq.~(\ref{eq-muSa0}) which, as already mentioned, derive from the ratio $\gamma/\beta=25/2$. Using $\gamma/\beta=25/2$, from Eq.\,(\ref{eq-muSa0}) 
we obtain $0.052 \approx - \frac{9}{2} \beta$, namely $\beta  \approx - 0.011$ and $\gamma \approx  - 0.144$,
in agreement with the left plot of Fig.\,\ref{fig-tbtg}.
So, shall  low energy SUSY be seen as a mirage induced by those particular ratios? 
Is it possible to find an $SU(5)$ model where those ratios are obtained from the breaking chains of the 24 and 75 irreps? 
We will come back to these questions in the following sections.

\end{itemize}

\subsection{Analysis of the contributions from $1$, $75$ and $200$}
\label{subsec-200}

\begin{figure}[htb!]
\vskip .5cm 
 \begin{center}
 \includegraphics[width=7.6 cm]{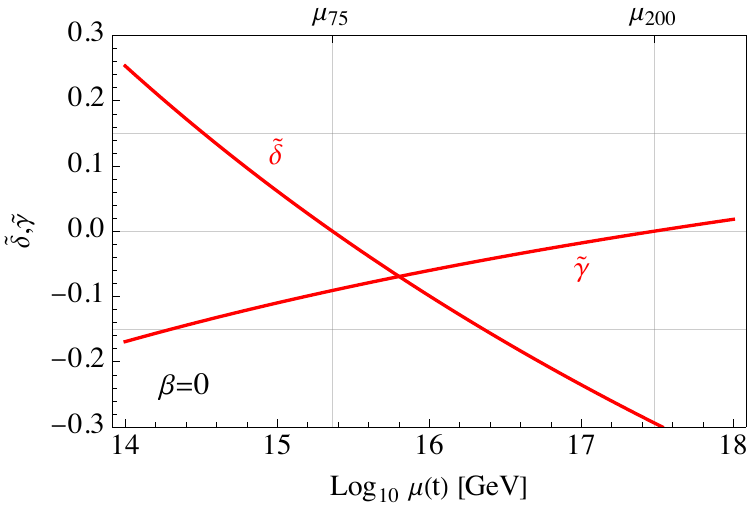} \,\,  \includegraphics[width=7.6 cm]{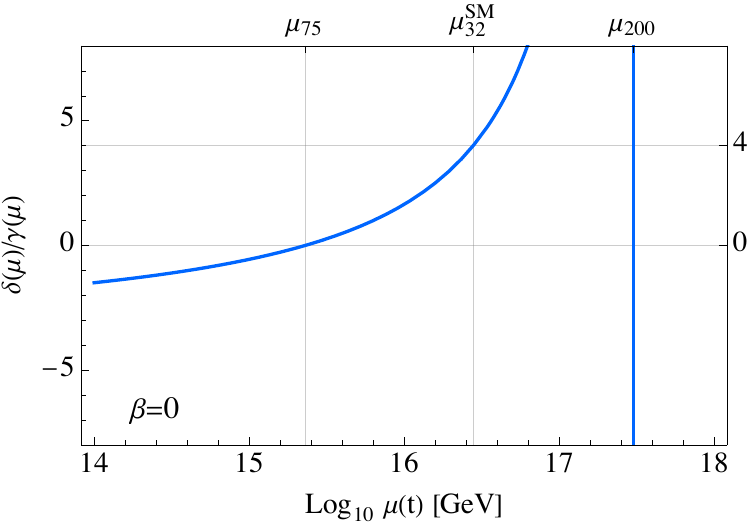} 
 \end{center}
\caption{\baselineskip=12 pt \small \it
Left panel: Solid lines are $\tilde \delta(\mu)$ and $\tilde \gamma(\mu)$ according to the exact solution, for $\tilde \beta=0$.
Right panel: Ratio $\delta(\mu)/\gamma(\mu)$.} 
\label{fig-tdtg}
\vskip .5 cm
\end{figure}

If we only allow for the presence of the representations $1, 75$ and $200$, 
the two unknowns $\tilde \gamma$ and $\tilde \delta$, can be univocally determined as a function of $\mu$, by using the exact solution (\ref{eq-tbtg-ex}) with $\tilde \beta=0$. 
The relevance of such curious scenario will be discussed in the following sections. The absence of the $24$ (the constraint $\tilde \beta=0$), is sufficient to determine that
\beq
\tilde \delta= -20 \frac{   f_{21}f_{31} +3 f_{31} -4 f_{21}}{ f_{21}f_{31}+15 f_{31} -40 f_{21}} \,\, , \,\, 
\tilde \gamma= -5 \frac{   f_{21}f_{31} - 9 f_{31} +8 f_{21}}{ f_{21}f_{31}+15 f_{31} -40 f_{21}} \,\,,
\eeq
which are shown in the left plot of Fig.~\ref{fig-tdtg}. 
Requiring that $|\tilde \delta|,|\tilde \gamma| \lesssim 0.15$, we see from the latter plot that GCU can be achieved for any scale $\mu$ in the range 
$10^{14.5} \lesssim \mu/{\rm{GeV}}\lesssim10^{16.4}$. 
The ratio $\delta/\gamma$ is shown in the right plot of Fig.~\ref{fig-tdtg}.

We now discuss a few interesting cases for GCU, taking $\alpha=0$ (the generalization to $\alpha \neq 0$ will be addressed in the following).

\begin{figure}[htb!]
\vskip .5cm 
 \begin{center}
  \includegraphics[width=7.6 cm]{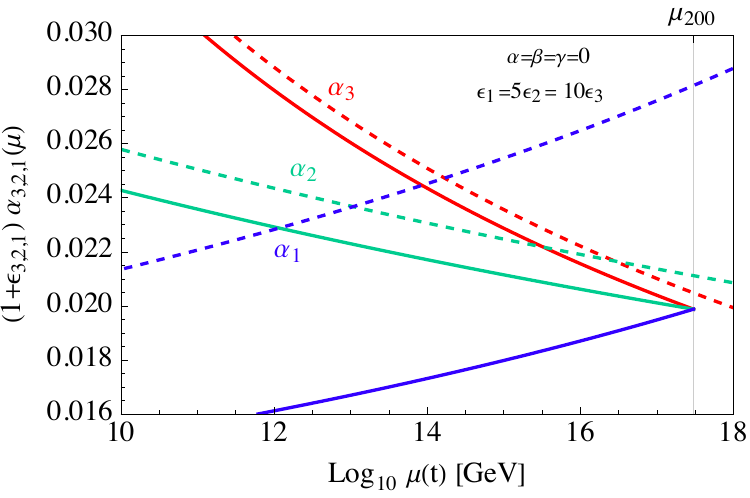}\,\,\,\,\,\,
 \includegraphics[width=7.6 cm]{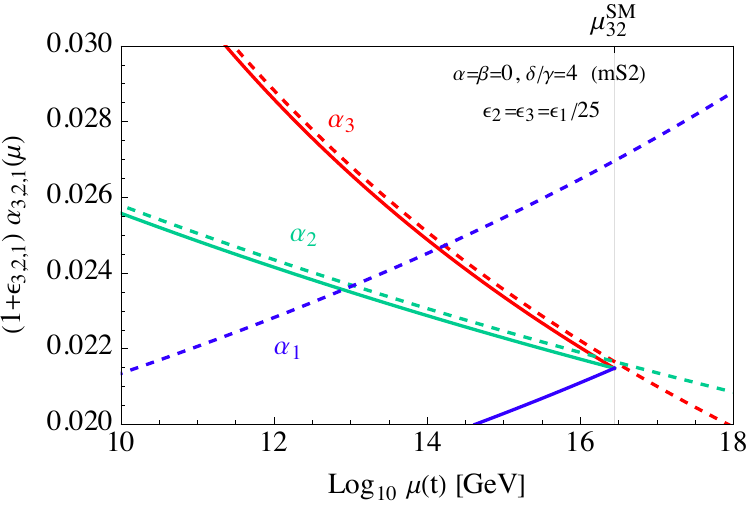}
 \end{center}
\caption{\baselineskip=12 pt \small \it
Left: GCU happens at $\mu_{200}$  by taking $-0.029\approx  \epsilon_1(\mu_{200}) \approx 5\epsilon_2(\mu_{200})  \approx 10 \epsilon_3(\mu_{200})$. 
Right: With $\alpha=\beta=0$ and $\delta/\gamma=4$, GCU happens at $\mu_{mS}$ by taking $0.052\approx \epsilon_2(\mu_{mS})\approx \epsilon_3(\mu_{mS}) \approx \epsilon_1(\mu_{mS})/25$.
Dashed lines are the gauge couplings in the SM. } 
\label{fig-unifb0}
\vskip .5 cm
\end{figure}

\begin{itemize}

\item 200 only ($\alpha=\beta=\gamma=0$): $M_X=\mu_{200} \approx 10^{17.5}\simeq 3.2\times 10^{17}$ GeV.\\
From our numerical analysis we found that (see also Ref.~\cite{Calmet:2009hp}), with $\tilde \delta \approx -0.29$ (a value which might jeopardize the validity of perturbation theory), 
GCU can be achieved without $24$ and $75$, at $\mu_{200} \approx 10^{17.5}$ GeV. In this case, taking $\alpha=0$, we have
\beq
-0.29\approx  \delta= \epsilon_1(\mu_{200}) \approx 5\, \epsilon_2(\mu_{200})  \approx 10\, \epsilon_3(\mu_{200}) \, .
\label{eq-mu200a0}
\eeq 
This scenario is shown in the left plot of Fig. \ref{fig-unifb0}. 

\item $mS_2$ ($\alpha=\beta=0$, $\delta/\gamma=4$): $M_X=\mu^{\rm SM}_{32}=\mu_{mS}=10^{16.4}$ GeV $\approx 2.8\times 10^{16}$ GeV. \\
Mirage SUSY can be realized in this scenario too, provided $\delta/\gamma=4$ (see the right plot of Fig.~\ref{fig-tdtg}), so that
\beq
-0.0081 \approx \frac{1}{25} \epsilon_1 (\mu_{mS})\approx \epsilon_2(\mu_{mS})\approx \epsilon_3(\mu_{mS}) \approx  \frac{1}{5} \gamma \,\, .
\label{eq-mSdelta}
\eeq
Within this scenario, that we are going to denote $mS_2$, the combinations $(1+\epsilon_s) \alpha_s(\mu)$ behave as shown in the right plot of Fig.~\ref{fig-unifb0}. 
Notice how it nearly looks like having a suitable normalization for the hypercharge, leading to GCU. 

\end{itemize}

\subsection{Comparison of the previous scenarios}

It has to be emphasized that the parameterization in Eq.~(\ref{eq-unif-gr}) precisely corresponds to the more general one proposed in Ref.\,\cite{Masina:2025klj} for the sake of studying how (partial or full) GCU might be achieved in new physics models;
in particular, partial unification of the non-Abelian couplings leads to a useful relation between $M_X$, $\epsilon_3$ and $\epsilon_2$, as
\beq
\frac{M_X} {\mu^{\rm SM}_{32}} = \, \exp \left( \frac{2 \pi}{\alpha^{\rm SM}_{32}}\, \frac{\epsilon_3 -\epsilon_2}{  (1+\epsilon_3)\, b^{\rm SM}_2 -(1+\epsilon_2)\, b^{\rm SM}_3  } \right) \,,
\label{eq-MX}
\eeq 
where $\mu_{32}^{\rm SM}\approx 2.8 \times 10^{16}$ GeV is the non-Abelian gauge couplings partial unification scale, 
$\alpha^{\rm SM}_{32}\equiv \alpha_2(\mu^{\rm SM}_{32})= \alpha_3(\mu^{\rm SM}_{32}) \approx 0.0217$, 
while $b_2^{\rm SM}= -19/6$ and $b_3^{\rm SM}=-7$ are the SM beta functions at one-loop. 
Notice that the magnitude of difference $\epsilon_3-\epsilon_2$ is related to the difference between $M_X$ and $\mu^{\rm SM}_{32}$. 

\begin{figure}[htb!]
\vskip .5cm 
 \begin{center}
 \includegraphics[width=7.8 cm]{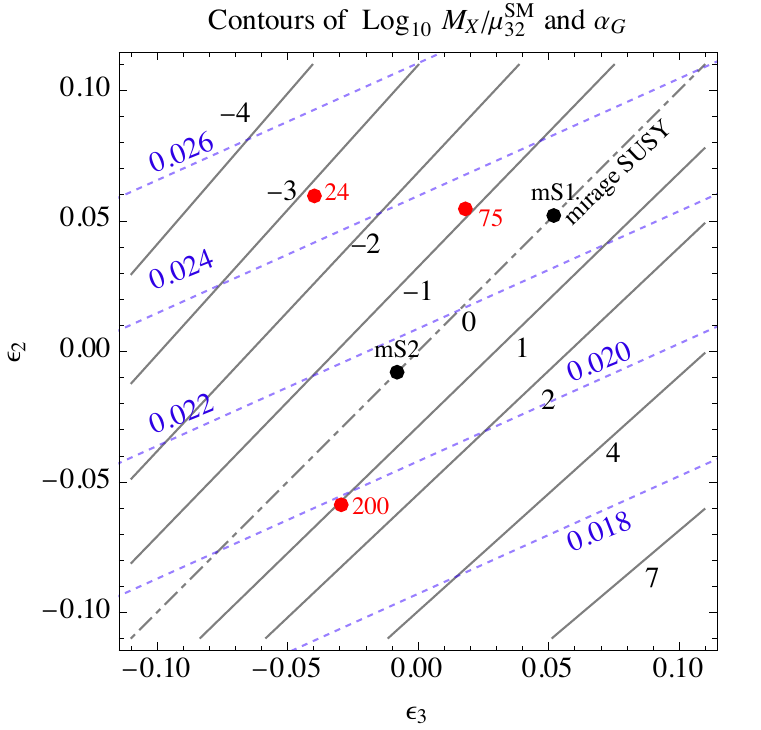}  
 \end{center}
\caption{\baselineskip=12 pt \small \it
Contours of ${\rm Log}_{10} M_X/\mu_{32}^{\rm SM}$ (solid black and dot-dashed black) and $\alpha_G$ (dashed blue). 
The circles emphasize the values corresponding to selected models with $\alpha=0$ discussed in the text.
} 
\label{fig-MXcont}
\vskip 0. cm
\end{figure}

It is interesting to localize the previous scenarios in the landscape of $\epsilon_2$ and $\epsilon_3$ (as was done in \cite{Masina:2025klj} for other models). 
Using Eq.\,(\ref{eq-MX}), this can be done by displaying the iso-levels of $\log_{10} M_X/\mu_{32}^{\rm SM}$ (solid lines) 
and $\alpha_G$ (dashed lines) in the plane $(\epsilon_2,\epsilon_3)$.
Fig.\,\ref{fig-MXcont} shows the location of the previous scenarios, and allows for a direct comparison with other models providing GCU, 
as those discussed in Ref.\,\cite{Masina:2025klj}. 

The scenarios denoted by $mS_1$ and $mS_2$ lie along the dot-dashed line, characterized by the relation $\epsilon_2=\epsilon_3$, 
where all mirage SUSY models are located.
Above the dot-dashed line, $\epsilon_3-\epsilon_2<0$, so that $M_X < \mu_{32}^{\rm SM}$; in this region we find the scenarios denoted as 24 only and 75 only,
for which $\epsilon_3-\epsilon_2\approx -0.1$ and $\epsilon_3-\epsilon_2 \approx-0.03$, respectively. 
Below the dot-dashed line, $\epsilon_3-\epsilon_2>0$, so that $M_X > \mu_{32}^{\rm SM}$; the 200 only scenario is found here, for which $\epsilon_3-\epsilon_2 \approx 0.03$.  

As already noticed, a non vanishing value of $\alpha$ does not change the scale where GCU takes place, 
as $\alpha$ acts as an offset for gauge couplings. 
Graphically, switching on $\alpha$ is equivalent to moving along the iso-levels of $M_X$ shown in Fig. \ref{fig-MXcont}: 
with $\alpha >0$ ($\alpha<0$), one moves towards larger (smaller) values of $\alpha_G$, hence to the top right (bottom left).
In the next section we will explore in more details the effect of varying $\alpha$.

\section{Detailed analysis of some relevant cases}
\label{sec-BU}

Let us fix $\mu=M_X$ as the scale where GCU happens via the corrections that we are studying.
We will relate the values of the $\epsilon_s(M_X)$ (hence of the parameters $\alpha, \beta, \gamma, \delta$) among themselves,
generalizing previous results derived considering $\alpha=0$. This is equivalent to include in Eq.\,(\ref{eq-opgr}) the singlet representation ($r=1$); 
even though such representation does not break $SU(5)$, it might in principle be present. 
The impact of the constraints from non-observation of proton decay will also be discussed.

\subsection{Generalization with $\alpha \neq 0$}

We now focus on the generalization of the three cases of Sec. \ref{subsec-75} with $\delta=0$ (in which GCU is achieved at $\mu_{24}$, $\mu_{75}$ and $\mu_{mS}$),
and of the two cases of Sec. \ref{subsec-200} with $\beta=0$ (in which GCU is achieved at $\mu_{200}$ and $\mu_{mS}$). The generalization consists in including the singlet irrep.

\begin{figure}[tb!]
\vskip 0.cm 
 \begin{center}
\includegraphics[width=7.8 cm]{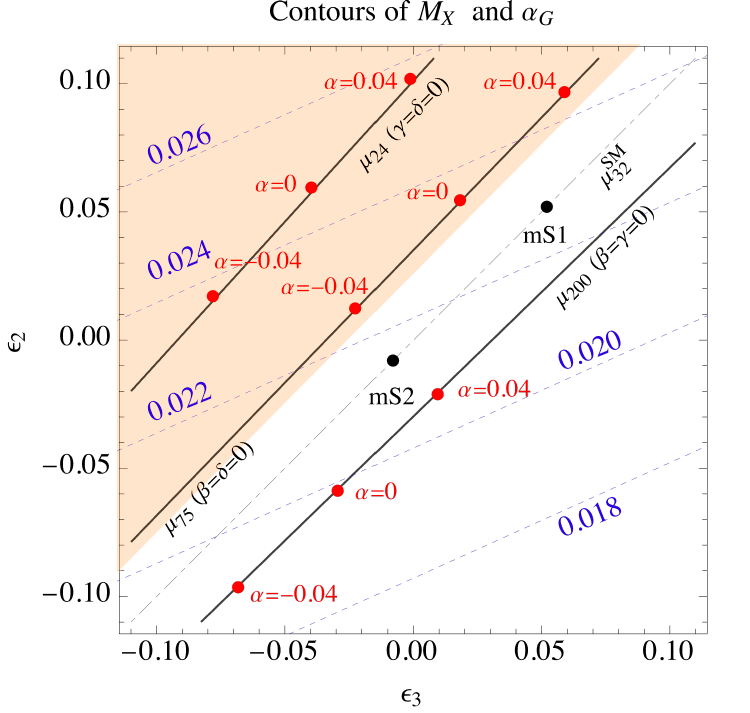}  
 \end{center}
\caption{\baselineskip=12 pt \small \it
Right: Contours of $M_X$ (solid black and dot-dashed black) and $\alpha_G$ (dashed blue). The effect of a non-vanishing value for $\alpha$ is shown. 
The shaded (orange) region is excluded by proton decay contraints.} 
\label{fig-MXcont-a}
\vskip .5 cm
\end{figure}

\begin{itemize}
\item $1+24$ only ($\gamma=\delta=0$): $M_X=\mu_{24}$\\
The generalization of Eq.~(\ref{eq-mu24a0}) to the case $\alpha\neq0$ is 
\beq
0.020 +\alpha \approx  \alpha+\beta = \epsilon_1(\mu_{24}) \approx \frac{1}{3} \epsilon_2(\mu_{24}) + \frac{2}{3} \alpha \approx - \frac{1}{2} \epsilon_3(\mu_{24}) + \frac{3}{2} \alpha \,\,.
\label{eq-1p24}
\eeq
As expected, there is no relation between $\alpha$ and $\beta$: 
the latter is fixed to about $0.02$, while the former is free and acts as an offset for GCU. 
The iso-level contour for $M_X=\mu_{24}$ corresponds to the line whose points have coordinates
$( \epsilon_2, \epsilon_3 ) \approx ( 0.06 , -0.04) +\alpha$, as shown in Fig.\,\ref{fig-MXcont-a}.
The three (red) dots emphasize the position associated to some particular values, $\alpha=0, \pm 0.04$;
hence, by increasing $\alpha$ one moves up-right along the line. Notice also that $\alpha_G \approx  0.0247 (1+ \alpha) $. 
\item $1+75$ only ($\beta=\delta=0$): $M_X=\mu_{75}$ \\
For $\alpha \neq 0$, Eq.\,(\ref{eq-mu75a0}) is generalized to
\beq
-0.091 +\alpha \approx  \alpha+\gamma = \epsilon_1(\mu_{75}) \approx - \frac{5}{3} \epsilon_2(\mu_{75}) + \frac{8}{3} \alpha \approx -5 \epsilon_3(\mu_{75}) + 6 \alpha \, .
\eeq
Again, there is no relation between $\gamma \approx -0.091$ and $\alpha$, which acts as an offset.
The iso-level contour corresponding to $M_X=\mu_{75}$ is the line $( \epsilon_2, \epsilon_3 ) \approx ( 0.05 , 0.02) +\alpha$, 
as shown in Fig.\,\ref{fig-MXcont-a}. In addition, we have $\alpha_G \approx  0.0235 (1+ \alpha)$.

\item$1+200$ only ($\beta=\gamma=0$): $M_X=\mu_{200}$ \\
Eq.~(\ref{eq-mu200a0}) is generalized to $\alpha \neq 0$ by
\beq
-0.3 + \alpha \approx  \delta + \alpha= \epsilon_1(\mu_{200}) \approx 5\, \epsilon_2(\mu_{200}) - 5   \alpha  \approx 10\, \epsilon_3(\mu_{200})- 10 \alpha \, .
\eeq
The line corresponding to these scenarios is $(\epsilon_2, \epsilon_3 ) \approx (- 0.06 ,- 0.03) +\alpha$, as shown in Fig.\,\ref{fig-MXcont-a}. In this case $\alpha_G \approx  0.020 (1+ \alpha)$.
\end{itemize}

We now focus on the two scenarios of the mirage SUSY type introduced in the previous section.
\begin{itemize}
\item $mS_1$ ($\delta=0$, $\gamma/\beta=25/2$): $M_X=\mu_{mS}$ \\
For $\alpha=0$, we derived Eq.~(\ref{eq-muSa0}) and found that $\gamma/\beta=25/2$ implies $\beta  \approx - 0.011$ and $\gamma \approx  - 0.144$.
For $\alpha \neq 0$, such relation is generalized to
\beq
0.052 + \alpha \approx \epsilon_2(\mu_{mS})\approx \epsilon_3(\mu_{mS}) \approx -\frac{1}{3}\epsilon_1(\mu_{mS})+ \frac{4}{3}\alpha \,\,,
\label{eq-gen-mS1}
\eeq
and the corresponding line in Fig. \ref{fig-MXcont-a} is simply $(\epsilon_2, \epsilon_3 ) \approx ( 0.052 , 0.052) +\alpha$.
Notice also that $\alpha_G \approx 0.0228 (1+\alpha)$.
\item $mS_2$ ($\beta=0$, $\delta/\gamma=4$): $M_X=\mu_{mS}$ \\
Eq.~(\ref{eq-mSdelta}) is generalized to $\alpha \neq 0$ by
\beq
-0.0081 + \alpha \approx \frac{1}{25} \epsilon_1 (\mu_{mS}) - \frac{1}{25} \alpha \approx \epsilon_2(\mu_{mS})\approx \epsilon_3(\mu_{mS}) \approx  \frac{1}{5} \gamma +\alpha \,\, .
\label{eq-gen-mS2}
\eeq
The corresponding line in Fig. \ref{fig-MXcont-a} is simply $(\epsilon_2, \epsilon_3 ) \approx ( -0.008 , -0.008) +\alpha$.
\end{itemize}

\subsection{Relation with proton decay}

A high value of $\mu=M_X$ for GCU is welcome to avoid problems with proton decay. 
As is well known, in $SU(5)$ GUTs, $d=6$ operators are induced by $X$ boson exchange. 
The main (non-SUSY) decay mode is $p \rightarrow e^+ \pi^0$, with a lifetime given by \cite{ParticleDataGroup:2024cfk}
(for the lattice coefficients see \cite{Yoo:2021gql}, and for analytic ones see Hisano et al.~\cite{Hisano:2012wq})
\beq
\tau_p = {\mathcal{O}}(1) \, \frac{  M_{X}^4}{\alpha_G^{2} m_p^5} \,\, ,
\eeq
where $m_p$ is the proton mass and $M_X$ is the $X$ boson mass, to be identified with the GCU scale.

So, to prolong $p$ lifetime, it would be better to have $\alpha_G$ as small as possible and $M_X$ as large as possible.
The present experimental bound from Super-Kamiokande is $\tau_p/{\rm{Br}}(p \rightarrow e^+ \pi^0)> 2.4 \times 10^{34}$ years at $90 \%$ CL,
which implies
\beq
M_{X} \gtrsim  \left( \frac{\alpha_G}{\alpha^{\rm SM}_{32}} \right)^{1/2} \,  4.5 \times 10^{15}  \, {\rm{GeV}}\,,
\label{eq-pdec-MX}
\eeq
where we recall that $\alpha^{\rm SM}_{32}=0.0217$. 

The bound is reported in Fig. \ref{fig-MXcont-a}: models in the shaded region are excluded.
This means that the previous cases denoted as $1+24$ only are severely ruled out, while those of the $1+75$ type are disfavored, 
since $\mu_{75} \approx 2.3 \times 10^{15}$ GeV.
The scenarios of the type 1$+200$ are instead fully acceptable, as well as mirage SUSY models.

Indeed, mirage SUSY turns out to be safe, since $M_X = \mu_{mS}=\mu^{\rm SM}_{32}=2.8 \times 10^{16}$ GeV, and the related value of $\alpha_G$ is quite small: $\alpha_G \approx 0.023 (1+\alpha)$ for $mS_1$ (and even smaller for $mS_2$).
On the other hand, in the case of low energy SUSY, one would have $\alpha_G \approx 0.038$ (and of course $M_X=\mu_{mS}$).
Hence, considering $mS_1$ for definiteness,  
proton lifetime for the decay $p \rightarrow e^+ \pi^0$ is longer with respect to the case of low energy SUSY:
the enhancement factor is $(0.038/0.023)^2 \approx 3$ in the case $\alpha=0$, and larger with $\alpha$ negative
\footnote{We recall however that in SUSY $SU(5)$ the calculation of the matrix element is slightly different than in the SM, see e.g.\,\cite{Hisano:2012wq}. 
In addition, the decay mode $p \rightarrow e^+ \pi^0$ is not even expected to be dominant. 
The main decay mode is expected to be $p \rightarrow K^+ \bar \nu$, arising from $d=5$ operators with heavy scalar triplet exchange.}.

\section{Relation with the doublet-triplet splitting problem}
\label{sec-DTS}

In this section, we investigate possible connections between GCU and the doublet-triplet splitting (DTS) problem.
In the following, we will distinguish two scenarios about the nature of the spontaneous $SU(5)$ breaking to the SM: \textit{i)} By propagating elementary scalars, as in the Higgs mechanism; 
\textit{ii)} By fermion condensates which dynamically break $SU(5)$ to the SM.

\subsection{Higgs breaking of $SU(5)$}

As already discussed in Sec.\,\ref{sec-GCU} (see also the explicit calculation leading to Eq.\,(\ref{eq-summ})), 
if we allow for the presence of scalar Higgses in the $SU(5)$ representations, $H_r=1, 24, 75, 200$, 
the non-renormalizable  $d=5$ operator of Eq.\,(\ref{eq-opgr})
induces the following corrections 
\bea
\epsilon_1 = \sum_{r} \epsilon_1^{(r)} =\alpha + \beta  +\gamma +\delta 
=   \frac{1}{\Lambda}   (c_1 v_1 -\frac{1}{2} \,c_{24} v_{24} - 5 \,c_{75} v_{75} +  10\, c_{200} v_{200} )  \, , \nonumber \\
\epsilon_2 = \sum_{r} \epsilon_2^{(r)} =\alpha + 3 \beta -\frac{3}{5} \gamma +\frac{1}{5} \delta 
=  \frac{1}{\Lambda} (c_1 v_1 -\frac{3}{2} \,c_{24} v_{24} + 3 \,c_{75}  v_{75}+  2\, c_{200} v_{200} ) \, ,\label{eq-general} \\
\epsilon_3 = \sum_r \epsilon_3^{(r)} =\alpha - 2 \beta  -\frac{1}{5} \gamma +\frac{1}{10} \delta 
=  \frac{1}{\Lambda}  (c_1 v_1 + c_{24} v_{24} + c_{75} v_{75}+  c_{200} v_{200} ) \, , \nonumber
\eea
where we recall the definitions 
$\langle H_1 \rangle =v_1 \,\mathbb 1_5$ and $\langle H_{24}\rangle=v_{24}\, {\rm diag}(\mathbb 1_3,-3/2\cdot \mathbb{1}_2)$,
while for more details about the contributions from $r=75,200$, we refer to App. \ref{app-SMmultiplets}.

As for the DTS problem (in a non-SUSY framework), let us consider the representation in which the SM Higgs doublet is contained, $H_5=(T,D)^T$. In general, triplet and doublet mass terms, denoted by $m_T$ and $m_D$ respectively, come from the following Lagrangian density terms
\bea
 {\mathcal L} \ni   
- {H}_{5}^\dagger [ m_5^2+m_1  \langle H_1  \rangle +m_{24} \langle H_{24} \rangle 
 + a_1 \langle H_1^2 \rangle  +a_{24} {\rm Tr} (\langle H_{24}  \rangle ^2)  +  b_{24} \langle H_{24} \rangle ^2 + c_{124} \langle H_{1}  H_{24} \rangle] H_5  &&
 \nonumber \\
 =- {T}^\dagger {m_T^2}  T   - {D}^\dagger {m_D^2}   D  \,,
 \,\,\,\,\,\,\,\,\,\,\,\,\,\,\,\,\,\, \,\,\,\,\,\,\,\,\,\,\,\,\,\,\,\,\,\, \,\,\,\,\,\,\,\,\,\,\,\,\,\,\,\,\,\, \,\,\,\,\,\,\,\,\,\,\,\,\,\,\,\,\,\, 
  \,\,\,\,\,\,\,\,\,\,\,\,\,\,\,\,\,\, \,\,\,\,\,\,\,\,\,\,\,\,\,\,\,\,\,\, \,\,\,\,\,\,\,\,\,\,\,\,\,\,\,\,\,\, \,\,\,\,\,\,\,\,\,\,\,\,\,\,\,\,\,\, \,\,\,\,\,\,\, \,\,\,\,\,\,\, \,\,\,\,\,\,\, \,\,\,\,\,\,\,\,\,\,
  \eea
 where the parameters $m_5, m_1, m_{24}$ have the dimension of a mass, while $a_1, a_{24}, b_{24}, c_{124}$ are dimensionless,
and we have
 \bea
&& m_T^2  = m^2_5 + m_1 v_1 + (m_{24}+c_{124} v_1) v_{24} + a_1 v_1^2 + \frac{15}{2} a_{24} v_{24}^2 + b_{24} v_{24}^2  \,\, ,\nonumber \\   
&& m_D^2 = m^2_5 + m_1 v_1 -\frac{3}{2} (m_{24}+c_{124} v_1) v_{24} + a_1 v_1^2 + \frac{15}{2} a_{24} v_{24}^2 + \frac{9}{4} b_{24} v_{24}^2    \,\, . \label{eq-mTmD}
 \eea
Even if present, the $75$ and $200$ would provide no contribution to doublet and triplet masses.
Notice also that the terms that can make a difference between the doublet and the triplet are those proportional to 
the combination $(m_{24}+c_{124} v_1)$ and $b_{24}$.
So, ensuring a vanishing mass to the doublet, while keeping the triplet at the GUT scale, can be seen as a problematic tuning.
In addition, even accepting the tuning, the doublet mass would be unstable against radiative corrections, unless invoking SUSY. 
Without SUSY, a reasonable possibility would be to promote the doublet to a pseudo-Nambu-Goldstone boson of some broken symmetry. 

So, by comparing Eqs.~(\ref{eq-general}) and (\ref{eq-mTmD}), the conclusion is that there is no direct link between GCU and DTS.
As we are now going to discuss, a link can however be established in the case that the representations are effective ones, 
sharing a common origin in a more fundamental theory.

Before doing this, let us discuss another possibility for solving the DTS problem, the so called Missing Partner Mechanism (MDM) \cite{Masiero:1982fe, Grinstein:1982um}.
It relies on the absence of $1,24$, and on the presence of the $75$ to break $SU(5)$. 
Adding scalars in $50, \overline{50}$ representations, it is possible to give mass to triplets, while keeping doublets (including the SM one) massless.
The MDM was originally proposed in the context of SUSY, where GCU is achieved, and was studied in some detail for instance in Refs.\,\cite{Altarelli:2000fu, Masina:2001pp}. 
The MDM can however be exploited also in a non-SUSY framework, as we are considering here; 
as for GCU, in this case one has to consider Eq.\,(\ref{eq-general}) with $\alpha=\beta=\delta=0$, 
which corresponds to scenario denoted by $75$ only (see the discussion in Sec. \ref{subsec-75} and Fig. \ref{fig-MXcont}).
So, also in the case of the non-SUSY MDM, one cannot find any direct link between the parameters involved in DTS and the parameter $\gamma$ involved in GCU.

\subsection{Dynamical breaking of $SU(5)$}
\label{subsec-dynbreak}

We will now consider a scenario where $SU(5)$ is broken dynamically. Previous studies where the GUT symmetry is broken dynamically by the presence of a strong dynamics have been performed mainly in the context of supersymmetric theories~\cite{Kugo:1994qr,Kitano:2001ie,Kitano:2006wm}. 
In the case of non-supersymmetric theories, an option is to start with a supersymmetric theory at high scales, and e.g.~consider a supergravity theory where local supersymmetry is spontaneously broken by a chiral superfield $Z$; by adopting e.g.~the Polonyi mechanism~\cite{Lahanas:1986uc}, 
the scale of supersymmetry breaking is $F_{Z}\sim M_X M_P$ leading to a gravitino mass $m_{3/2}\sim M_X$. 
In this way the theory for scales below $M_X$ would behave as a non-supersymmetric theory. A study in that direction is postponed to further investigations. 

Here we will just assume that there is a confining group $G$ which becomes strong at an IR scale $\Lambda_G\simeq M_X$. All conventional GUT fields are then singlets under $G$, i.e.~the matter fermions $10_F$ and $\bar 5_F$, the Higgs boson $5_H$ and the gauge bosons $24_G$. Moreover, we will assume that the UV theory for scales $\mu\gtrsim M_X$ contains a set of Dirac (anomaly free) fermions $F_{R}$ (and $\bar F_{R}$) which transform under the $R=5,10,24$ (and $\bar R=\bar 5, \overline{ 10},24$) representations of $SU(5)$ and under the $R_G$ (and $\bar R_G$) representation of $G$. In this paper we will be agnostic about the group $G$ and the representation $R_G$ and $\bar R_G$, but just will consider the most attractive channel $X$ corresponding to the maximal binding strength $\kappa(X)=2C_2(R_G)-C_2(X)$ to be the singlet representation of $G$, $X=1$, as $C_2(1)=0$, such that the bilinear 
$\bar F_{R} \times F_{R}\equiv \mathcal{T}_R$ (in the singlet representation of the confining group $G$) produces the condensate as $\langle \mathcal{T}_R \rangle \simeq \Lambda_G^3$~\cite{Raby:1979my}. 
The theory is then a QCD-like, or better a technicolor-like (techniGUT) theory. For some more recent references on binding strength, see e.g.~Refs.~\cite{Bolognesi:2021jzs,Bai:2021tgl}.

In this way, we recognize that the fermion condensates generate the effective representations of Eq.~(\ref{eq-opgr})
\beq
\mathcal{T}_R=\bar F_R \times F_ R \supset 1+24+a_{R} \cdot 75+b_{R} \cdot 200 \,,
\label{eq:descomposicion} 
\eeq
where the constants $a_R$ and $b_R$ take the following values~\footnote{The parameters $a_R$ and $b_R$ introduced here should not be confused with those introduced in Eq.\,(\ref{eq-mTmD}).}: $a_{5}=b_{5}=0$; $a_{10}=1, b_{10}=0$; and $a_{24}=b_{24}=1$. The confining dynamics alone produces a condensate but is blind to alignment. Gauge interactions through the effective potential $V_{\rm eff}(\mathcal{T_R})$ will break the vacua degeneracy leading to the vacuum alignment characteristic of non-supersymmetric theories. In this paper we will consider 
the VEVs along the different components of (\ref{eq:descomposicion}) as free parameters, and study their role for GCU and the DTS mechanism. 
In the case of supersymmetric theories the vacuum alignment is much simpler, as it relies on the search for flat directions, and will be done elsewhere.

Taking $R=5$ is the most economical possibility\,\footnote{This is in a sense a "really-minimal $SU(5)$", as there are just 10 fermionic fields, instead of the $24$ scalar fields of the "minimal $SU(5)$" model.}; the representations 1 and 24 are obtained as effective ones, while the 75 and the 200 are absent. 
Taking instead $R=10$, the 200 representation is not obtained.
Finally, putting the fermion in the adjoint representation, $R=24$, one gets all the representations that might contribute to the non-renormalizable operator of Eq.\,(\ref{eq-opgr}).

We will assume that the condensates break $SU(5)$ along the direction of $SU(3)\otimes SU(2)\otimes U(1)$, 
such that 
\beq
\frac{\langle\mathcal{T}_R \rangle}{\Lambda_G^2}= \frac{\langle \bar F_{R} \times F_{R} \rangle}{ \Lambda_G^2} 
=  \langle H^{(R)}_{1} \rangle + \langle H^{(R)}_{24} \rangle + a_{R}\, \langle H^{(R)}_{75} \rangle + b_{R} \langle H^{(R)}_{200} \rangle  \,\, ,
\label{eq-RRcond}
\eeq
the effective Higgses $H^{(R)}_r$, acquire VEVs as discussed in App. \ref{app-SMmultiplets}.

The branching rules of $SU(5)$ under $SU(3)\otimes SU(2)\otimes U(1)$ are such that $\bar F_R \times F_R \supset (3,2)_{5/6}+(\bar 3,2)_{-5/6}$, which are the 12 Goldstone bosons~\footnote{For instance for $R=5$, $\bar F_{\alpha}\gamma_5 F^a$ and $F^\alpha \gamma_5 \bar F_{a}$, where $\alpha=1,2,3$ are $SU(3)$ indices (color), and $a=1,2$ $SU(2)$ ones.} eaten by the $X_\mu$ and $Y_\mu$ gauge fields to become massive. 
Moreover in the heavy spectrum, above the condensation scale, there are mesons, from other components of $(\bar F_{R}\times F_{R})/\Lambda_G^2$, 
and baryons, from e.g.~$B_{R}=(\epsilon_{ABCDE}F_{R}^A F_{R}^B F_{R}^C F_{R}^D F_{R}^E)/\Lambda_G^6$, which are singlets under $SU(5)$.

\subsubsection{The gauge coupling unification}

Let us first focus on GCU. At the condensation scale $\mu\sim \Lambda_G$, we find (see App.~\ref{app:kinetic}) the non-renormalizable  $d=5$ operator 
\beq
  -\frac{1}{4}  \frac{c_R}{\Lambda_G} {\rm Tr}(G_{\mu\nu} \, (\langle  H^{(R)}_{1} \rangle + \langle  H^{(R)}_{24} \rangle + a_R\langle  H^{(R)}_{75} \rangle+b_R \langle  H^{(R)}_{200} \rangle) \,G^{\mu\nu} )\,\, ,
  \label{eq:kinetic}
\eeq
where $R$ can be chosen among $5,10,24$. The crucial point is that, unlike Eq.~(\ref{eq-opgr}), the dynamical breaking case of Eq.~(\ref{eq:kinetic})
leads to a common coefficient, $c_R$, for the contributions of the various effective representations. 
As a result, the relative magnitudes of the contributions to the $\epsilon$'s just depend on the effective VEVs. 

Defining $\langle H^{(R)}_r \rangle$ as discussed in App.~\ref{app-SMmultiplets}, one obtains Eq.~(\ref{eq-summ}),
from which it can be seen that the above operator induces the following corrections
\bea
\epsilon_1 =\sum_r  \epsilon_1^{(r)}  =\alpha + \beta  + a_R\, \gamma + b_R\, \delta 
=   \frac{c_R}{\Lambda_G}   ( v_1 -\frac{1}{2}  v_{24} - 5 \,a_R\, v_{75}+  10 \,b_R\, v_{200} )  \,\, , \nonumber \\
\epsilon_2 = \sum_r \epsilon_2^{(r)} =\alpha + 3 \,\beta -\frac{3}{5} a_R \,\gamma + \frac{1}{5} b_R \,\delta 
=  \frac{c_R}{\Lambda_G} (v_1 -\frac{3}{2} v_{24} + 3\, a_R\,  v_{75}+  2\, b_R\, v_{200}) \,\, , \nonumber \\
\epsilon_3 = \sum_r \epsilon_3^{(r)} = \alpha - 2 \,\beta  -\frac{1}{5} a_R\, \gamma +\frac{1}{10} b_R \,\delta 
=  \frac{c_R}{\Lambda_G}  (v_1 + v_{24} + a_R\, v_{75}+  b_R\, v_{200}) \, .\,\, 
\label{eq-epsilons-FF}
\eea
One thus obtains relations among $\alpha, \beta,\gamma$ and $\delta$, which depend only on the relations among the various VEVs, $v_r$ .

\subsubsection{The doublet-triplet splitting}

One expects that also the DTS condition will depend on the effectives VEVs, and it will thus be possible to relate GCU and DTS. In this section we present the main ideas, 
and will analyze in detail the phenomenology and the model building, related to 
the various possible fermion representations, $R=5,10,24$, in the next section. 

We will consider an effective Lagrangian as (see App.~\ref{app:DTS} for a particular simple model)
\beq
\mathcal L_{\rm eff}=-\frac{\lambda_R^2}{\Lambda_G} H^\dagger_5 ( \bar F_{R} \times F_R) H_5 ,
\eeq
giving rise, upon fermion condensation as in Eq.\,(\ref{eq-RRcond}), to the mass terms

\beq
\mathcal L_{\rm eff}=-\lambda_R^2\,\Lambda_G H_5^\dagger \left(\langle H_{1}^{(R)}+H_{24}^{(R)}\rangle\right)H_5\,. 
\label{eq:DTS}
\eeq

Defining $\langle H^{(R)}_r \rangle$ and the associated VEVs, $v^{(R)}_r$, as discussed in App.~\ref{app-SMmultiplets}, 
it is possible to write
\beq
\langle H^{(R)}_1 \rangle =v_1 \,\mathbb{1}_5 \,\, , \,\, \langle H^{(R)}_{24}\rangle=v_{24}\, {\rm diag}(\mathbb{1}_3,-3/2\cdot \mathbb{1}_2) \,
\eeq
for any $R=5,10,24$.
The operator above then splits the doublet and triplet masses as
\beq
 {\mathcal L}_{\rm eff} \ni - {T^\dagger}  \underbrace{ \lambda_R^2 \Lambda_G \left(v_1 + v_{24}\right) }_{m_T^2}  T -
  {D^\dagger}  \underbrace{ \lambda_R^2 \Lambda_G \left(v_1 -\frac{3}{2} v_{24}\right)}_{m_D^2}   D \,\, .
  \label{eq-eff-lag}
 \eeq
Notice that there is no contribution to DTS from the $H_{75}^{(R)}$ and $H_{200}^{(R)}$ of Eq.\,(\ref{eq-RRcond}), even in case they are present in the fermion condensate, as they do not couple to $\bar 5\times 5$.
 
The doublet is massless if the condition 
\beq
m_D^2=\lambda_R^2 \Lambda_G \left(v_1 -\frac{3}{2} v_{24} \right)=0 
\label{eq-dtcond}
\eeq
is satisfied, in which case the triplet mass is 
$m_T^2 =\lambda_R^2 \Lambda_G ( v_1 + v_{24})= \frac{5}{2} \lambda_R^2 \Lambda_G  v_{24}$.

\subsubsection{Doublet-triplet splitting constraints on gauge coupling unification}

By comparing Eq.\,(\ref{eq-dtcond}) with Eq.~(\ref{eq-epsilons-FF}), one can recognize that the triplet mass has the structure of the "1+24" contribution to $\epsilon_3$,
while the doublet mass has the structure of the "1+24" contribution to $\epsilon_2$.
The DTS condition thus implies 
\beq
\alpha= -3 \beta \,\, .
\eeq
If we rewrite Eq.~(\ref{eq-epsilons-FF}) using $\alpha= -3 \beta$, and 
the equivalent relation $v_1 =\frac{3}{2} v_{24}$, we obtain
\bea
\epsilon_1 = -2 \, \beta  +a_R\,\gamma +b_R\, \delta 
=   \frac{c_R}{\Lambda_G}   (  v_{24} - 5\, a_R\, v_{75}+  10 \,b_R\, v_{200} )  \,\, , \nonumber \\
\epsilon_2 =  -\frac{3}{5} a_R \,\gamma + \frac{1}{5} b_R\,\delta 
=  \frac{c_R}{\Lambda_G} ( 3 \,a_R\,  v_{75}+  2\, b_R\, v_{200}) \,\, , \nonumber \\
\epsilon_3 = - 5 \, \beta  -\frac{1}{5} a_R\, \gamma +\frac{1}{10} b_R \, \delta 
=  \frac{c_R}{\Lambda_G}  \left(\frac{5}{2} v_{24} + a_R\, v_{75}+  b_R\, v_{200}\right) \,\, .
\label{eq-epsilons-FF-bis}
\eea
Using the above expressions in the system (\ref{eq-unif-gr-2}) accounting for GCU (where we substitute the condition $\alpha =- 3 \beta$), 
we obtain the equations
\bea
&&\beta (\mu) = \frac{ -4 (2+ b_R\,\delta) f_{21}  + (2+ b_R\,\delta/10)  f_{21} f_{31} + (6+3  b_R\,\delta/2)  f_{31}   }{3 (9 f_{31} +f_{21} (-2+5 f_{31}))} \,\,\, , \,\,\, \nonumber \\ 
 &&a_R\,  \gamma (\mu) =\frac{ -3 (5+8 \, b_R\,\delta) f_{31} +(5+ b_R\,\delta) f_{21} (-2+5 f_{31} )  }{3 (9 f_{31} +f_{21} (-2+5 f_{31}))} \,\,.
\label{eq-bg-RR}
\eea

\begin{figure}[htb!]
\vskip .5cm 
 \begin{center}
 \includegraphics[width=7 cm]{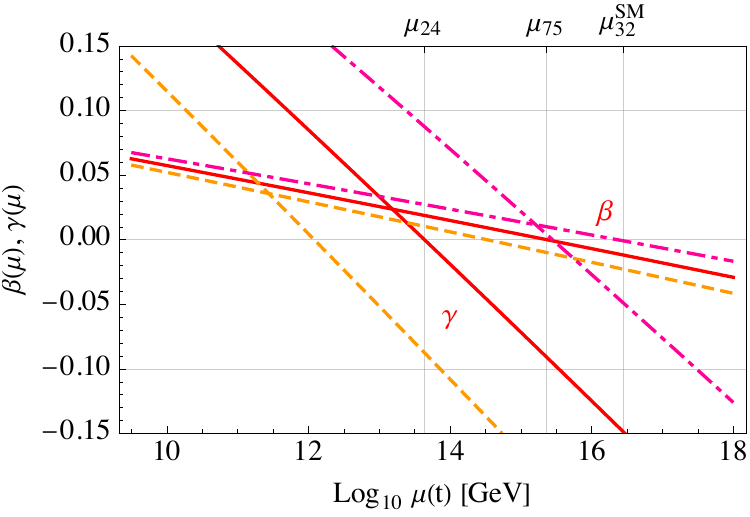}  \,\, 
  \includegraphics[width=7 cm]{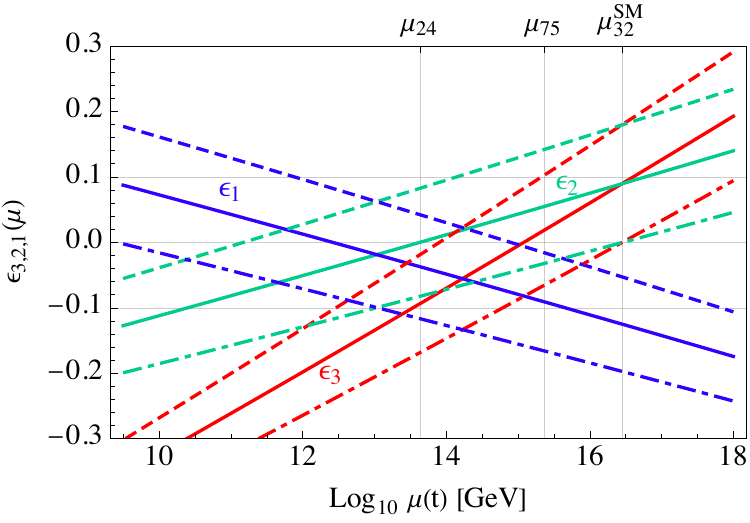}  \,\, 
 \end{center}
\caption{\baselineskip=12 pt \small \it
Left: The solutions for $\beta$ and $\gamma$ in Eq.~(\ref{eq-bg-RR}), with $\delta=0$ (solid), $\delta=0.15$ (dashed) and $\delta=-0.15$ (dot-dashed). 
Right:  The associated $\epsilon$ corrections, according to Eq.~(\ref{eq-epsilons-FF-bis}).} 
\label{fig-RR}
\vskip .5 cm
\end{figure}

Let us focus on the most general case $R=24$, so that the parameters $a_R, b_R$ in Eq.~(\ref{eq-bg-RR}) are equal to one.
For fixed values of $\delta$, the dependence of $\beta(\mu)$ and $\gamma(\mu)$ is shown in the left plot of Fig.\,\ref{fig-RR}, 
while the associated $\epsilon$ corrections are shown in the right plot; the solid, dashed and dot-dashed lines refer respectively to $\delta=0,0.15,-0.15$.
In the case $R=10$, $a_R$ in Eq.~(\ref{eq-bg-RR}) is equal to one,  while $b_R$ vanishes: this case is thus described by the solid curve ($\delta=0$) in Fig.\,\ref{fig-RR}.
In the case $R=5$, both $a_R$ and $b_R$ are vanishing; in particular the second equation in~(\ref{eq-bg-RR}), gives a constraint on $\mu$, which is satisfied only for 
$\mu=\mu_{24}$, as can be seen from the dependence of $\gamma(\mu)$ in the left plot of Fig.\,\ref{fig-RR}.

\section{The phenomenology of dynamical GUT breaking}
\label{sec-PD}

In this section we will consider in turn GCU and DTS for the cases of dynamical breaking such that $\langle \bar F_{R} \times F_R \rangle \neq 0$, 
for the three different cases where the condensates correspond to the representations $R=5,10,24$; the latter will be also respectively denoted by, $F$ (Five), $T$ (Ten), $A$ (Adjoint), 
in order to (hopefully) use a more effective notation. 

\subsection{Fermion condensates from $R=5$}

We introduce the fermion five-plets $\bar F \equiv \bar F_5$ and $F \equiv  F_5$, 
so that the tensor 
\beq
\frac{{\mathcal {T}}_A^B}{\Lambda_G^2} \equiv \frac{\bar F_A \times F^B}{\Lambda_G^2} =( {H_{1}^{(5)}}+{H_{24}^{(5)}} )_A^B =S_A^B + \Sigma_A^B \,,
\eeq
with indices $A,B=(1,\dots ,5)$, can be decomposed into the singlet ($S$) and adjoint ($\Sigma$) representations respectively.
We recall that $\langle S_A^B\rangle=v_1 \delta_A^B$ and $\langle\Sigma_A^B\rangle=v_{24} (\mathbb{1}_3,-3/2\cdot \mathbb{1}_2)$.

We denote the Higgs five-plets by ${\bar H}_{\bar 5}=(\bar T,\bar D)$ and $H_5=(T,D)^T$;
as we have seen in the previous section, the induced effective Lagrangian after fermion condensation is given by Eq.\,(\ref{eq-eff-lag}).
The doublet is massless if the condition (\ref{eq-dtcond}) is satisfied, 
in which case the triplet square mass is $m_T^2 = 5 \lambda_F^2 \Lambda_G v_{24}/2$,
where we have introduced the notation $\lambda_F \equiv \lambda_5$.

We now discuss in more detail the possible origin of $v_1$ and $v_{24}$. Recall that $v_1\neq 0$ does not break $SU(5)$, while $v_{24}\neq 0$ does. 
From the analysis in App.~\ref{app-sing-24}, it turns out that
\beq
v_1= \frac{1}{5} (\sqrt{3} s^{(5)}_1 + \sqrt{2} s^{(5)}_2) \,\,\, , \,\,\, v_{24}= \frac{\sqrt{2} }{\sqrt{3}\cdot 5} (\sqrt{2} s^{(5)}_1 - \sqrt{3} s^{(5)}_2)  \,\,
\eeq
where 
\beq
s^{(5)}_1= \frac{1 }{\sqrt{3} } \frac{ \langle {\mathcal{T}}_1^1 +{\mathcal{T}}_2^2 +{\mathcal{T}}_3^3 \rangle}{\Lambda_G^2}  \,\,\, , \,\,\, 
s^{(5)}_2= \frac{1 }{\sqrt{2} } \frac{ \langle {\mathcal{T}}_4^4 +{\mathcal{T}}_5^5 \rangle }{\Lambda_G^2}\,\,\, ,
\eeq
which can be non-vanishing as they are SM singlet VEVs. 
Notice that
\beq
m_T^2
= \frac{\lambda_F^2 \Lambda_G s^{(5)}_1}{\sqrt{3}} \,\,\, , \,\,\, 
m_D^2
=\frac{\lambda_F^2\Lambda_G s^{(5)}_2}{\sqrt{2}} \,\, .
\eeq
The triplet mass squared is thus proportional to $s^{(5)}_1$, while the DTS condition (\ref{eq-dtcond}) is equivalent to $s^{(5)}_2=0$.  
The breaking of $SU(5)$ to the SM and the DTS are simultaneously achieved if there is a mechanism such that
the SM color singlet $\langle {\mathcal{T}}_1^1 +{\mathcal{T}}_2^2 +{\mathcal{T}}_3^3 \rangle \neq 0$, 
while the SM weak isospin singlet $\langle {\mathcal{T}}_4^4 +{\mathcal{T}}_5^5 \rangle=0$. 

In Ref.~\cite{Kitano:2001ie} it was proven, in the context of supersymmetric theories, that the condition $s_2^{(5)}=0$ is a flat-direction of the supersymmetric potential and thus a supersymmetric minimum, which triggers automatically the DTS mechanism without any fine-tuning. One could here argue along similar lines, provided that supersymmetry be spontaneously broken at a scale $\sim \Lambda_G$. This is an interesting avenue although 
outside the scope of the present paper. Still, for the non-supersymmetric GUT we are here considering it is a very predictive scenario that we will be exploring.

As for the GCU, the correction to the kinetic term Lagrangian, Eq.\,(\ref{eq:kinetic}), becomes simply
\beq
\delta {\mathcal L} =-\frac{1}{4} 
 \frac{c_F}{\Lambda_G}  {\rm Tr}( G_{\mu\nu} \, \langle S + \Sigma \rangle\,G^{\mu\nu} )\,,
\label{eq-F}
\eeq
where $c_F \equiv c_5$ and, consistently with Eq.\,(\ref{eq-epsilons-FF}), one obtains
\bea
\epsilon_1 &=& \epsilon_1^{(1)} +\epsilon_1^{(24)} =   \alpha + \beta  =\frac{c_F}{\Lambda_G}\left(v_1 - \frac{1}{2} v_{24} \right )
=  \frac{c_F}{\Lambda_G}   \left(\frac{2}{5} \frac{s^{(5)}_1}{\sqrt{3}} + \frac{3}{5} \frac{s^{(5)}_2}{\sqrt{2}} \right) \, \nonumber \\
\epsilon_2 &=& \epsilon_2^{(1)} +\epsilon_2^{(24)} = \alpha + 3 \beta  = \frac{c_F}{\Lambda_G}\left( v_1 -\frac{3}{2} v_{24}\right )
=  \frac{c_F}{\Lambda_G}  \frac{s^{(5)}_2}{\sqrt{2}} =  c_F \frac{ m_D^2}{ \lambda_5^2 \Lambda_G^2} \,\, ,\label{eq-eps-FFF}  \\
\epsilon_3 &=& \epsilon_3^{(1)} +\epsilon_3^{(24)} = \alpha - 2 \beta  = \frac{c_F}{\Lambda_G} (v_1 + v_{24})  
=  \frac{c_F}{\Lambda_G}  \frac{s^{(5)}_1}{\sqrt{3}} = c_F\frac{ m_T^2}{\lambda_5^2 \Lambda_G^2}  \,\, \nonumber  \,.
\eea
As we can see from Eq.~(\ref{eq-eps-FFF}), the contributions leading to GCU are directly related to the DTS mechanism. 
In particular, the condition (\ref{eq-dtcond}) applied to (\ref{eq-eps-FFF}) implies that $\epsilon_2=0$, or equivalently 
$\alpha =- 3 \beta$. Summarizing, we obtain $\epsilon_1=-2\beta$, $\epsilon_2=0$, $\epsilon_3= -5\beta$,
so that $\epsilon_3/\epsilon_1=5/2$.

\begin{figure}[htb!]
\vskip .5cm 
 \begin{center}
  \includegraphics[width=7 cm]{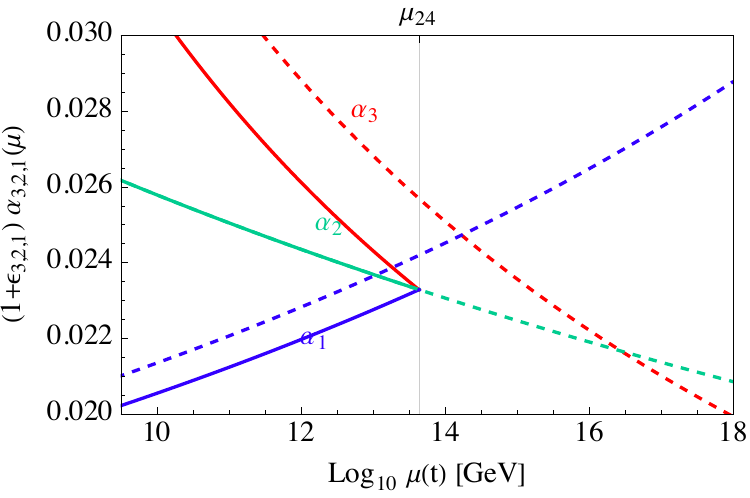}  
 \end{center}
\caption{\baselineskip=12 pt \small \it
GCU when the fermion condensate ${\overline 5} \times 5$ solves the DTS problem. } 
\label{fig-FF}
\vskip .5 cm
\end{figure}

We now carry out the phenomenological analysis for this scenario, initiated with the discussion leading to Eq.~(\ref{eq-bg-RR}).
We recognize that this case corresponds to a particular case of Eq.~(\ref{eq-1p24}), 
so that $M_X=\mu_{24}$, $\beta \approx 0.020$ and $\alpha \approx -  0.056$, together with $\epsilon_1 \approx -0.04, \epsilon_2=0, \epsilon_3 \approx - 0.09$.
The related quantities $(1+\epsilon_s)\alpha_s(\mu)$, with $s=1,2,3$, are explicitly shown in Fig.~\ref{fig-FF}.

Notice that this scenario corresponds to a single point in Fig.\,\ref{fig-MXcont-DT}, 
the one with coordinates $(\epsilon_3,\epsilon_2)=(- 0.09,0)$, explicitly indicated by means of the (green) star lying on the iso-level contour of $M_X=\mu_{24}$.
As for proton decay, 
we have a quite small value for $\alpha_G \approx 0.0247 (1+\alpha) \approx 0.023$; 
despite this, the scale $M_X$ is too low, and the constraint on the proton decay, Eq.\,(\ref{eq-pdec-MX}), is not satisfied;
the (green) star indeed falls in the excluded shaded region.

This minimalistic and elegant scenario, where DTS and GCU are univocally related, is thus not viable. 
We envisage two main roads to overcome this impasse:
i) introducing condensates generating also higher dimensional representations, like the $75$ and the $200$, as we are going to discuss next;
ii) introducing more five-plets, and possibly relating them to flavor; this path will not be followed here.

\subsection{Fermion condensates from $R=10$}

We introduce the fermion ten-plets $\bar T \equiv \bar F_{10}$ and $T= F_{10}$, in the $\overline{10}$ and $10$ representation respectively, so that we can 
exploit a 10 by 10 matrix notation and decompose the tensor
\beq
\frac{ {\mathcal{T}}_{CD}^{AB}}{\Lambda_G^2} \equiv \frac{ \bar T_{CD} \times T^{AB} } {\Lambda_G^2}=( {H_{1}^{(10)}}+{H_{24}^{(10)}} +{H_{75}^{(10)}})_{CD}^{AB}= S_{CD}^{AB} + \Sigma_{CD}^{AB}+ Y_{CD}^{AB} \,,
\eeq
where 
the latter effective representations are the singlet, adjoint and 75 respectively, whose form is derived in App.~\ref{app-75}. 
The ordering of the matrix elements is such that entries from 1 to 3 are related to color, and the last entry is related to the weak hypercharge.
The breaking to the SM is realized for
 \bea
\langle S\rangle=v_1\cdot {\rm diag}(\mathbb{1}_3,\mathbb{1}_6,1) \,\,\, , \,\,\, 
\langle\Sigma\rangle=6 \,v_{24}\cdot {\rm diag}\left(\mathbb{1}_3,-\frac{1}{4}\cdot \mathbb{1}_6,\frac{3}{2}\right) \,,\nn \\
\langle Y\rangle=-3\, v_{75}\cdot {\rm diag}(\mathbb{1}_3,- \mathbb{1}_6, 3) \,, \,\,\,\,\,\,\,\,\,\,\,\,\,\,\,\,\,\,\,\,\,\,\,\,\,\,\,\,\,\,\,\,\,\,\,\,\,\,\,\,\,\,\,
\eea
where 
\bea
& &v_1 =\frac{1}{10} (\sqrt{3} s^{(10)}_1 +\sqrt{6} s^{(10)}_2 +s^{(10)}_3) \,,\nn \\
&&v_{24}=\frac{\sqrt{2}}{45\sqrt{3}} (2 \sqrt{2} s^{(10)}_1 -s^{(10)}_2 -\sqrt{6} s^{(10)}_3) \,, \nn\\
&&v_{75}=-\frac{1}{18\sqrt{3} } ( s^{(10)}_1 - \sqrt{2} s^{(10)}_2 +\sqrt{3} s^{(10)}_3) \,,
\label{eq-10-v}
\eea
and the explicit form of the three SM singlets, $s^{(10)}_i$ with $i=1,2,3$, is given in App.~\ref{app-sing-75}.

\begin{figure}[tb!]
\vskip 0.cm 
 \begin{center}
\includegraphics[width=9 cm]{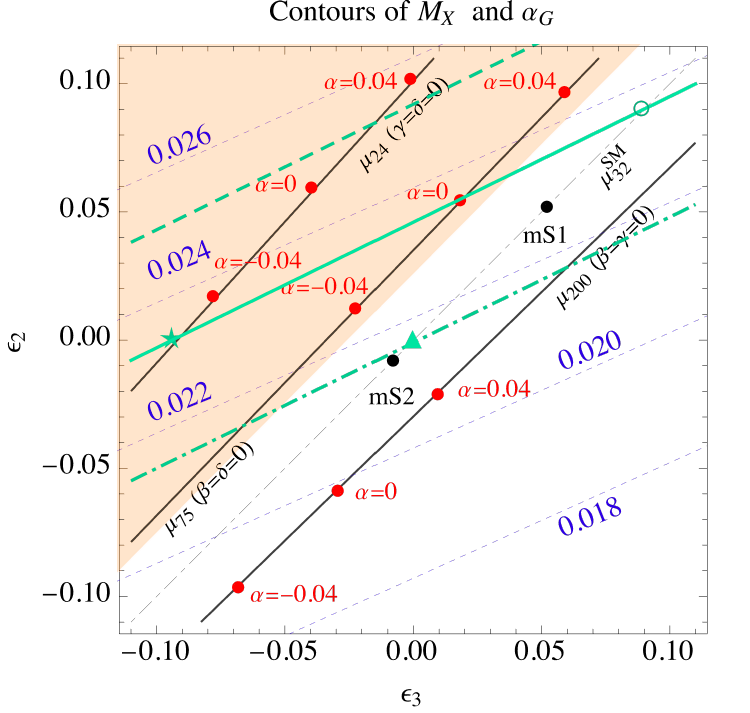}  
 \end{center}
\caption{\baselineskip=12 pt \small \it
Contours of $M_X$ (solid black and dot-dashed black) and $\alpha_G$ (dashed blue). 
Configurations fulfilling the DTS condition $\alpha=-3\beta$ are shown in green. For $R=5$, the DTS condition is satisfied by a single point, corresponding to the green star on top of the iso-level $M_X=\mu_{24}$. For $R=10$, the DTS condition is satisfied within the green solid line; the open circle on top of $\mu_{32}^{\rm SM}$ represents the case $mS_1$, for which $\gamma/\beta=25/2$. For $R=24$, the DTS condition is satisfied in the whole plane: the dashed and dot-dashed green lines represent the models with 
$\delta=+0.15$ and $\delta=-0.15$; the latter crosses the iso-level $\mu_{32}^{\rm SM}$ at the point represented by the green triangle, for which $\epsilon_2=\epsilon_3=0$ and $\alpha_G=\alpha_{32}^{\rm SM}$.   
The shaded (orange) region is excluded by proton decay constraints.} 
\label{fig-MXcont-DT}
\vskip .5 cm
\end{figure}

The non-renormalizable operator relevant to GCU, given in Eq.\,(\ref{eq:kinetic}), is now
\beq
\delta {\mathcal L} 
=-\frac{1}{4} \frac{c_{T}}{\Lambda_G}  {\rm Tr}( G_{\mu\nu} \, \langle S + \Sigma +Y\rangle\,G^{\mu\nu} )
\label{eq-T}
\eeq
where $c_T \equiv c_{10}$ and, consistently with Eq.\,(\ref{eq-epsilons-FF}), one has
\bea
\epsilon_1 &=& \epsilon_1^{(1)} +\epsilon_1^{(24)} +\epsilon_1^{(75)} = \alpha + \beta  + \gamma
= \frac{c_{T}}{\Lambda_G} (v_1 - \frac{1}{2} v_{24} -5 v_{75}  )\,,     \nonumber\\
\epsilon_2 &=& \epsilon_2^{(1)} +\epsilon_2^{(24)} +\epsilon_2^{(75)}  =  \alpha + 3 \beta  -\frac{3}{5} \gamma 
=\frac{c_{T}}{\Lambda_G} (v_1 -\frac{3}{2} v_{24}+3  v_{75} ) \,,  \nonumber \\
\epsilon_3 &=& \epsilon_3^{(1)} +\epsilon_3^{(24)} +\epsilon_3^{(75)} = \alpha - 2 \beta -\frac{1}{5} \gamma 
= \frac{c_{T}}{\Lambda_G} (v_1 + v_{24}+ v_{75}  )  \,. 
\label{eq-eps-TT}
\eea
It can be shown that the contribution to $\epsilon_2$ is proportional to
\beq
v_1 -\frac{3}{2} v_{24} +3 v_{75} =\frac{1}{\sqrt{6}} s^{(10)}_2 \,,
\label{eq-TT-eps2}
\eeq
where $s^{(10)}_2$ turns out to be a SM singlet combination for both color and weak isospin charges.

The DTS for the Higgs $H_5=(T,D)^T$ works  {\it mutatis mutandis} as in the previous case, 
as the 75 does not contribute to the induced effective Lagrangian after fermion condensation, Eq.\,(\ref{eq-eff-lag}).
Hence, the DTS condition is 
 \beq
0= v_1 -\frac{3}{2} v_{24}  =\frac{1}{6\sqrt{3}} (s^{(10)}_1 +2\sqrt{2} s^{(10)}_2 +\sqrt{3} s^{(10)}_3) \,.
\eeq
The DTS condition above applied to Eq.\,(\ref{eq-eps-TT}) implies that $\alpha =- 3 \beta$, while applied to Eq.\,(\ref{eq-10-v})
allows to write the VEVs as a function of $s^{(10)}_2$ and $s^{(10)}_3$ only,
\beq
v_1 = - \frac{\sqrt{6}  }{10 } s^{(10)}_2 \left(  1 + \frac{ \sqrt{2}} { \sqrt{3}} \frac{s^{(10)}_3}{s^{(10)}_2} \right) \, , \,\,
v_{24}= -\frac{\sqrt{6}}{ 15} s^{(10)}_2  \left( 1+ \frac{\sqrt{2}}{\sqrt{3} }  \frac{s^{(10)}_3}{s^{(10)}_2}\right) \, , \,\,
v_{75}=  \frac{\sqrt{6} }{18 }  s^{(10)}_2   \,. 
\eeq
We thus recognize that the ratio $v_{75}/v_{24}$ depends on the ratio $s^{(10)}_3/s^{(10)}_2$,
\beq
\frac{v_{75}}{v_{24}} = -\frac{ 5 }{6 } \frac{1}{ 1+ \frac{\sqrt{2}}{\sqrt{3}}  \frac{s^{(10)}_3}{s^{(10)}_2}} \,.
\label{eq-T-vevs}
\eeq

We now carry out the phenomenological analysis for this scenario. As anticipated, we obtain Eq.~(\ref{eq-bg-RR}) with the constraint that $\delta=0$.
As shown by the solid lines in Fig.~\ref{fig-RR},
unification can be achieved for any desired scale $\mu$, hence even larger than $\mu_{24}$ (which instead was the only scale allowed in the case $R=5$ previously studied).

The solid green line of Fig. \ref{fig-MXcont-DT} explicitly shows the location of these models solving simultaneously the DTS and GCU; 
such line corresponds to the function $\epsilon_2= \epsilon_3 + 0.45$.
Notice that this line includes the (green) star on top of the iso-level for $M_X=\mu_{24}$ (where $\epsilon_2=\gamma=0$), 
as well as the $\alpha=0$ (red) dot of the iso-level for $M_X=\mu_{75}$ 
(in the case $\alpha=0$, due to the DTS condition, $\beta=0$ too, also implying that the triplet should acquire mass from another mechanism, as for instance the MDM). 
Both these scenarios are however in trouble with proton decay constraints.
On the other hand, all points along the (green solid) branch which do not fall into the shaded (orange) region are safe with respect to proton decay constraints. 
So, there are many possible viable scenarios with the effective $10 \times \overline{10}$, where the conditions on DTS and GCU are enforced.
\begin{figure}[htb!]
\vskip .5cm 
 \begin{center}
 \includegraphics[width=7 cm]{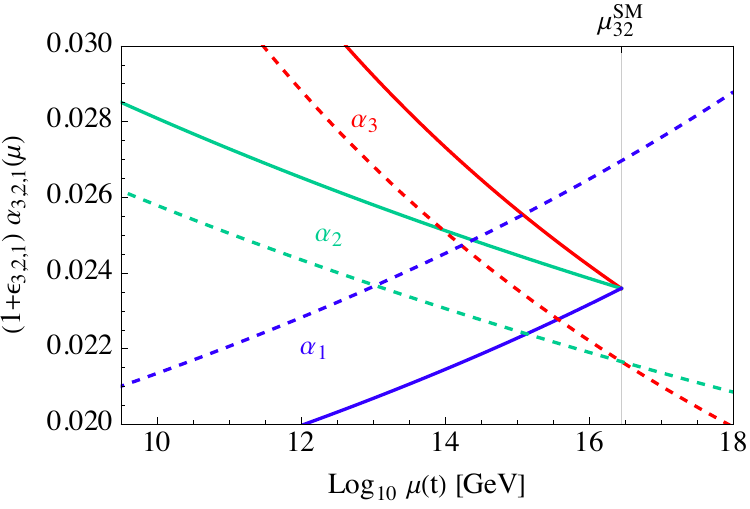}  
 \end{center}
\caption{\baselineskip=12 pt \small \it
GCU for the particular case $mS_1$,
when the fermion condensate ${\overline {10}} \times 10$ solves the DTS problem. } 
\label{fig-TT}
\vskip .5 cm
\end{figure}

In the following, we focus our attention on a particular case, that is the point on top of the iso-level with $M_X=\mu_{32}^{\rm SM}$,
emphasized by the open green circle in Fig.\,\ref{fig-MXcont-DT}. 
It corresponds to the mirage SUSY case denoted by $mS_1$ (for which $\gamma/\beta=25/2$), and described by Eq.\,(\ref{eq-gen-mS1}), 
with the additional constraint  $\alpha = -3 \beta \approx  0.03$ (see also the solid lines of Fig. \ref{fig-RR}), so that $\epsilon_2=\epsilon_3\approx 0.09$.
As shown in Fig.~\ref{fig-TT}, unification is achieved with $\alpha_G \approx 0.0236$ (which is significantly smaller than in the low energy SUSY case).
As a consequence, this scenario satisfies the constraints on proton lifetime, Eq.\,(\ref{eq-pdec-MX}).  
Notice also that, using the DTS condition ($v_1=\frac{3}{2}v_{24}$) and the condition $\gamma/\beta=25/2$, from Eq.\,(\ref{eq-eps-TT}) 
we get the corresponding values of the parameters
\beq
\alpha=\frac{3}{2}\frac{c_{T}}{\Lambda_G}v_{24},\quad \beta=-\frac{1}{2}\frac{c_{T}}{\Lambda_G}v_{24},\quad \gamma=- \frac{25}{4}\frac{c_{T}}{\Lambda_G}\,v_{24} \,.
\eeq
and the relation between the VEVs
\beq
v_{75} =\frac{5}{4}v_{24} \,  ,
\eeq
which, according to Eq. (\ref{eq-T-vevs}), implies that $mS_1$ is associated to the following specific direction implementing the dynamical breaking
\beq
 \frac{ 3 }{  \sqrt{6}} \frac{< {\mathcal{T}}^{\alpha b}_{\alpha b}>}   {<{\mathcal{T}}_{ab}^{ab}>} =    \frac{s^{(10)}_3}{s^{(10)}_2}  = -  \frac{5}{ \sqrt{6} }  \,\, .
\eeq

\subsection{Fermion condensates for $R=24$}

We finally introduce the fermion 24-plet fields $\bar A\equiv \bar F_{24}$ and $A \equiv F_{24}$, so that the tensor has the decomposition
\beq
\frac{{\mathcal{T}}_{CD}^{AB}}{\Lambda_G^2} \equiv  \frac{\bar A_{CD} \times A^{AB}}{\Lambda_G^2} 
= ( {H_{1}^{(24)}}+{H_{24}^{(24)}} +{H_{75}^{(24)}} +{H_{200}^{(24)}})_{CD}^{AB}
=  S_{CD}^{AB} + \Sigma_{CD}^{AB}+ Y_{CD}^{AB}+ Z_{CD}^{AB}\,,
\eeq
where the latter effective representations are the singlet, adjoint, 75 and 200 respectively. 
Their form has been derived explicitly in Ref.~\cite{Calmet:2009hp}, by using $24\times 24$ matrices. 
The ordering of the elements is such that entries from 1 to 8 are related to color, those from 9 to 11 to weak isospin, the 12th entry is related to the hypercharge, 
and those from 13 to 24 are related to the heavy leptoquarks.
The breaking to the SM is realized when
 \bea
\langle S\rangle=v_1 \, \mathbb{1}_{24} \,\,\, , \,\,\, 
\langle\Sigma\rangle= v_{24} \, {\rm diag}(\mathbb{1}_8,-\frac{3}{2}\cdot\mathbb{1}_3,-\frac{1}{2}, -\frac{1}{4}\cdot \mathbb{1}_{12}) \,, \nonumber  \\
\langle Y\rangle = v_{75}\, {\rm diag}(\mathbb{1}_8, 3\cdot \mathbb{1}_3, -5,- \mathbb{1}_{12}) \,\,\,, \,\,\, 
\langle Z\rangle= v_{200}\, {\rm diag}(\mathbb{1}_8, 2\cdot \mathbb{1}_3, 10,-2\cdot\mathbb{1}_{12})  \,\, .
\eea

The non-renormalizable operator of Eq.\,(\ref{eq:kinetic}) now includes all the contributions which, according to the previous notation, read
\beq
\delta {\mathcal L} = -\frac{1}{4}\frac{c_{A}}{\Lambda_G}  {\rm Tr}( G_{\mu\nu} \, \langle S + \Sigma +Y +Z \rangle\,G^{\mu\nu} ) 
\label{eq-A}
\eeq
where $c_A \equiv c_{24}$ and, consistently with Eq.\,(\ref{eq-epsilons-FF}), one has
\bea
\epsilon_1 &=& \epsilon_1^{(1)} +\epsilon_1^{(24)} +\epsilon_1^{(75)}+\epsilon_1^{(200)}=\alpha + \beta  + \gamma + \delta
=  \frac{c_{A}}{\Lambda_G} (v_1 - \frac{1}{2} v_{24} - 5 v_{75} + 10 v_{200}  )\,\,  , \nonumber\\
\epsilon_2 &=& \epsilon_2^{(1)} +\epsilon_2^{(24)} +\epsilon_2^{(75)} +\epsilon_2^{(200)} =\alpha + 3 \beta  -\frac{3}{5} \gamma + \frac{1}{5} \delta
=  \frac{c_{A}}{\Lambda_G} (v_1 -\frac{3}{2} v_{24}+3  v_{75} + 2 v_{200} ) \,\, ,\nonumber \\
\epsilon_3 &=& \epsilon_3^{(1)} +\epsilon_3^{(24)} +\epsilon_3^{(75)} +\epsilon_3^{(200)} = \alpha - 2 \beta -\frac{1}{5} \gamma + \frac{1}{10} \delta
=  \frac{c_{A}}{\Lambda_G} (v_1 + v_{24}+v_{75} +v_{200}   )   \,. 
\label{eq-eps-AA}
\eea

The DTS works as in the previous cases, since the 75 and 200 do not contribute to the effective Lagrangian after fermion condensation, Eq.\,(\ref{eq-eff-lag}).
In particular, the DTS condition, $v_1 -\frac{3}{2} v_{24}  = 0$, always implies that $\alpha =- 3 \beta$. 
Also in this scenario, where we have one more parameter with respect to the case with $R=10$, 
GCU and DTS can be achieved at any scale, $\mu$. 

In order to carry out the phenomenological analysis for this scenario, we come back to Eq.\,(\ref{eq-bg-RR}) and Fig.\,\ref{fig-RR}. 
As for the location of these models in Fig.\,\ref{fig-MXcont-DT}, we obtain an area, rather than a line; 
as an example, the dashed and dot-dashed green lines correspond to selected values of $\delta$, respectively $\delta=+0.15$ and 
$\delta=-0.15$.
As a result, all the mirage SUSY models can be reproduced within a condensate with $R=24$. 

\begin{figure}[htb!]
\vskip .5cm 
 \begin{center}
\includegraphics[width=7 cm]{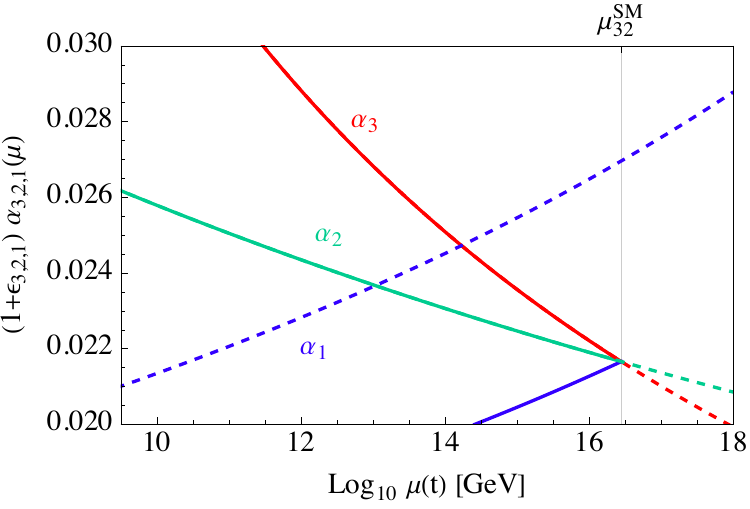}  
 \end{center}
\caption{\baselineskip=12 pt \small \it
GCU for the particular mirage SUSY case in which $\alpha_G=\alpha^{\rm SM}_{32}$, when the fermion condensate $24\times 24$ solves the DTS. } 
\label{fig-AA}
\vskip .5 cm
\end{figure}

Let us focus in particular on the scenario in which DTS is realized, and GCU is achieved at $M_X=\mu_{32}^{\rm SM}$, with the additional condition that
$\alpha_G=\alpha^{\rm SM}_{32}$. The latter condition implies $\epsilon_2=\epsilon_3=0$, hence $\delta=3 \gamma$ and $\beta=\gamma/50$, which leads to
\beq
-\alpha= 3 \beta= \frac{3}{50} \gamma = \frac{1}{50} \delta \,.
\eeq
All the job of unification is thus carried out by $\epsilon_1$, 
\beq
-0.197 \approx \epsilon_1= \alpha + \beta  + \gamma + \delta = 198 \,\beta \,,
\eeq
as shown in Fig. \ref{fig-AA}. Numerically, it turns out that $\beta \simeq 0.001$; it is interesting that such a small value leads to successful GCU and DTS within this scenario. On the other hand, 
notice that we have $\delta \approx -0.15$. The green triangle of Fig.\,\ref{fig-MXcont-DT} displays the location of this particular scenario.

Notice that implementing the DTS condition ($\alpha=-3\beta$) within the scenario $mS_2$ (for which $\delta/\gamma=4$ and $\beta=0$), leads to $\alpha=\beta=0$,
that is to a vanishing mass for the Higgs triplet. This scenario, corresponding to the black circle labelled by $mS_2$ in Fig.\,\ref{fig-MXcont-DT}, would thus require an alternative mechanism for the triplet to acquire a mass, as for instance the MDM. 

\section{Conclusions}
\label{sec-conc}

We considered the phenomenology of a non-supersymmetric $SU(5)$ framework~\cite{Georgi:1974sy} in which GCU is achieved because the SM gauge couplings receive UV origin corrections,
denoted by $\epsilon_i$ ($i=1,2,3$), which are induced by a non-renormalizable  $d=5$ kinetic operator~\cite{Hill:1983xh, Shafi:1983gz, Panagiotakopoulos:1984wf, Hall:1992kq, Calmet:2008df, Bhatt:2008qb, Chakrabortty:2008zk, Calmet:2009hp}. 
The representations that can in principle be involved are denoted by $H_r$, with $r=1,24,75,200$, and $SU(5)$ is broken to the SM by the VEVs of the representations $r=24,75,200$.

We first studied the impact on GCU, especially on the unification scale $M_X$ and the unified coupling $\alpha_G$, for the various representations, separately and in some relevant combinations. According to the proposal in Ref. \cite{Masina:2025klj}, we compared the most representative scenarios by mapping them in the plane of the non-Abelian corrections, $\epsilon_2$ and $\epsilon_3$, as done in Figs.~\ref{fig-MXcont} and \ref{fig-MXcont-a}.

We then questioned the nature of the representations $H_r$, in relation with the doublet-triplet splitting problem. The electroweak scale Higgs doublet and its conjugate are indeed part of the $5$ and $\bar 5$ scalar Higgs representations, together with the GUT scale triplet and anti-triplet.     
In the former literature, $H_r$ have been considered to be elementary scalars, spontaneously breaking $SU(5)$ (for $r>1$) as in the Higgs mechanism;
in this case there is a huge number of parameters in the Lagrangian, and there is no direct relation between the parameters providing GCU and those involved in the DTS.   

A reduction in the number of parameters is achieved by postulating that the representations $H_r$ are effective ones, coming from some UV completion of $SU(5)$.
In particular, we postulate that they are originated from the condensation of extra Dirac fermions, $F_R$ and $\bar F_R$, sitting in the representation $R$ and $\bar R$ of $SU(5)$, and in the representation $R_G$ and $\bar R_G$ of some confining group $G$.
Interestingly, the phenomenology of this framework can be studied without specifying the details of the group $G$ and its matter content.
Indeed, in the case that $R=5, 10, 24$, the breaking of $SU(5)$ is achieved dynamically by meson in the effective theory having the form $\langle \bar F_R \times F_R\rangle$,  
which are singlets under $G$ and whose VEV is given by a specific combination of the $\langle H_r^{(R)}\rangle$, 
which differs upon the choice of $R$, according to Eq.~(\ref{eq-RRcond}).
We showed that, due to the reduction in the number of parameters, in this framework it is possible to relate the parameters involved in the DTS with those accounting for GCU; the relation is univocally determined for $R=5$, and relaxes increasingly for $R=10$ and $R=24$.   

Specifically, solving simultaneously the DTS and GCU in the case $R=5$, 
leads to a sharp prediction for $M_X$, which turns out to be too low and in conflict with proton decay constraints;
this scenario is represented by the green star in Fig.\,\ref{fig-MXcont-DT}. 
In the case $R=10$, the parameter space solving simultaneously DTS and GCU becomes a line, as shown via the solid green one in Fig.\,\ref{fig-MXcont-DT}; 
now $M_X$ can be made large enough to escape the proton decay bounds. 
With $R=24$, the whole area of Fig.\,\ref{fig-MXcont-DT} is at disposal to solve the DTS and GCU, with even larger possibilities to meet proton decay constraints.
  
Notice that, in the above framework, we have considered the Higgs doublet as an elementary scalar which is contained into the Higgs five-plet $H_5$ of $SU(5)$. Of course the possibility of considering the Higgs five-plet as a fermion condensate does exist in case there exist extra fermions in the 5 and 10 irreps $F$ and $T$ (and $\bar F$, $\bar T$) and condensates as $\langle T^{AB}\otimes \bar F_{B}\rangle=H_5^A$ and $\langle \bar T_{AB}\otimes F^B\rangle=H_{5A}^\dagger$. The DTS coupling can be generated as in the case of elementatry scalar Higgs five-plets, as described in the App.~\ref{app:DTS}.

Summarizing, we proved that in the case of a dynamical breaking of the GUT group -- here in particular we considered $SU(5)$ -- it is possible to achieve gauge coupling unification and simultaneously solve the doublet-triplet splitting problem. We find it remarkable that the phenomenological implications of such a scenario can be grasped 
even without entering into the details of the confining group. In particular, constraints from proton decay imply that the extra fermions which are going to condensate, dynamically breaking $SU(5)$, cannot belong to the $5$ representation (the fundamental representation) of $SU(5)$;
interestingly, viable models require such fermions to belong to larger representations, such as the $10$ or $24$.

To conclude, let us provide some comments about possible extensions of the present work, to be explored elsewhere.
An interesting possibility to build viable models is to consider more than a single $5$ representation, thus "flavoring" the fermion condensates; in this way it might be possible to achieve GCU and DTS simultaneously, avoiding constraints on proton decay. 
Notice also that in the present work the Higgs was put in the $5$ representation of $SU(5)$, as usually done; another interesting possibility would be to put it in the $45$ representation, and to assess the phenomenological impact of such choice.

\section*{\large Acknowledgments}

IM acknowledges partial support by the research project TAsP (Theoretical Astroparticle Physics) funded by the Istituto Nazionale di Fisica Nucleare (INFN). 
The work of MQ is also supported by the grant PID2023-146686NB-C31 funded by
MICIU/AEI/10.13039/501100011033/ and by FEDER, EU. IFAE is partially funded by
the CERCA program of the Generalitat de Catalunya. MQ acknowledges financial support from the Spanish Ministry of Science and Innovation (MICINN) through the Spanish State Research Agency, under Severo Ochoa Centres of Excellence Programme 2025-2029 (CEX2024001442-S).

\appendix
\vskip 1.cm

\appendix
\numberwithin{equation}{section}

\section{Group Theory for $SU(5)$}
\label{appgroup}

In this appendix we summarize the methods adopted to deal with tensors in $SU(5)$. 
Large representations (like the $75$) can be conveniently written as products of smaller representations (like $10$ and $\overline{10}$). 
In this way, it is not difficult to recognize how the SM multiplets are embedded in the large representations.

\subsection{Basic Conventions for the Tensors in $SU(5)$}

In $SU(5)$ the matter fields of a single family are contained in the $10$ and $\overline{5}$ representations.
We write them following the conventions of \cite{Masina:2001pp}  
(see also the Particle Data Group review on {\it Grand Unified Theories}\,\cite{ParticleDataGroup:2024cfk}), so that
\beq
\psi_{10}=\frac{1}{\sqrt{2}}
\left( \begin{matrix}  0&   u^c_3& -u^c_2  & u^1&d^1 \cr
            -u^c_3   &   0        &  u^c_1  & u^2 &  d^2 \cr
           u^c_2  &   -u^c_1 &       0     &   u^3  &  d^3 \cr
            -u^1      &   -u^2    &  -u^3     &     0    &   e^c \cr
            -d^1     &     -d^2    &   -d^3    &  -e^c   &   0     \end{matrix}
\right)~,
\,\,\, \, {\psi}_{\bar 5} = \left( \begin{matrix} d^c_1 \cr d^c_2 \cr d^c_3 \cr e \cr -\nu_e  \end{matrix} \right)~~.
\label{psi5bar}
\eeq
Our conventions are such that the fundamental representation corresponds to a superscript in tensorial
notation and the antifundamental (conjugated fundamental) to a subscript in tensorial notation.
The $\overline{5}$ representation,  ${\psi_{\bar 5}}_A$, is written as a tensor with a single lower index, $A=1,..,5$;
the $10$ representation, $\psi_{10}^{AB}$, is written as a tensor with two antisymmetric upper indices, $A,B=1,..,5$.
Note that the $SU(3)$ indices correspond to $A,B=1,2,3$, 
while the $SU(2)$ ones are selected for $A,B=4,5$. 
We will sometimes write $\alpha$ ($a$) instead of $A$ if $A=1,2,3$ ($4,5$). 

As for scalars, the SM Higgs doublet fits in the fundamental representation, $H_5$,
\beq
H_5 = \left( \begin{matrix} T^1 \cr T^2 \cr T^3  \cr D^+ \cr D^0 \end{matrix} \right)~~~~,
\eeq
breaking $SU(2)$ when $\langle D^0 \rangle = v/\sqrt{2}$.
In a non-supersymmetric context~\footnote{In a supersymmetric framework instead, one has to introduce two Higgs fields, $ {\bar H}_{\bar 5\, A} $ and $H_5^A$, 
belonging to the $\overline{5}$ and $5$ representations respectively, 
\beq
{\bar H}_{\bar 5} = \left( \begin{matrix} 
 {T_{d}}_1 \cr {T_{d}}_2 \cr{T_{d}}_3  \cr D_{d}^- \cr -D_{d}^0 \end{matrix} \right)~~,~~
H_5 = \left( \begin{matrix} T_{u}^1 \cr T_{u}^2 \cr T_{u}^3  \cr D_{u}^+ \cr D_{u}^0 
\end{matrix} \right)~~~~,
\eeq
containing the SM Higgs doublet and its conjugate. When $SU(2)$ is spontaneously broken, $\langle D_{d}^0 \rangle = v_d/\sqrt{2}, \langle D_{u}^0 \rangle = v_u/\sqrt{2}$, giving rise to down and up quark masses respectively.}, 
one can introduce just $H_5^A$ and its conjugate, $(H_5)^\dagger_A$, which transforms as a $\bar 5$.   
	    
We now discuss how to write large representations in terms of smaller ones.

\subsubsection{Effective representations from $\bar 5 \times 5$}

We define $\overline{5}_{A}  \times  5^{B}\equiv T^{B}_{A}$, which can be split according to
\beq
T^{B}_{A}=\underbrace{S^{B}_{A}}_{1}+\underbrace{\Sigma^{B}_{A}}_{24} 
\label{tab}
\eeq
where we identify the $1$ with $S$, the $24$ with $\Sigma$. 
Here and in the following summation over repeated indices has to be understood. 
The traceless condition for the 24 is thus $\Sigma^{A}_{A}=0$.

Defining
\beq
s\equiv T_{A}^{A}~~~,
\label{eq-defs}
\eeq 
one has
\beq
S^{B}_{A}=\frac{1}{5} \delta^B_A \,s ~~,
\,\,\, \Sigma^{B}_{A}= T_A^B -\frac{1}{5} \delta^B_A \,s  \,\, .
\label{sing-adj-ab}
\eeq

\subsubsection{Effective representations from $\overline{10} \times 10$}
\label{app-75}

We define $10^{AB} \times\overline{10}_{CD} \equiv T^{A B}_{C D}$,
so that $T^{A B}_{C D}$ is antisymmetric in the upper and lower pair
of indices and thus corresponds to 100 independent fields.
It can be split according to
\beq
T^{A B}_{C D}=\underbrace{S^{A B}_{C D}}_{1}+\underbrace{\Sigma^{A B}_{C D}}_{24}+
\underbrace{Y^{A B}_{C D}}_{75}
\label{tabcd}
\eeq
where we identify the $1$ with $S$, the $24$ with $\Sigma$ and the $75$ with $Y$.

Defining
\beq
s\equiv T_{MN}^{MN}~~~,\,\,\,\, \sigma^A_C \equiv T^{MA}_{MC} -\frac{1}{5} \delta^A_C T^{MN}_{MN}~~
\eeq 
one has
\beq
S^{AB}_{CD}=\frac{1}{20} (\delta^A_C \delta^B_D -\delta^A_D \delta^B_C) T^{M N}_{M N } ~~,
\label{sab}
\eeq

\bea
\Sigma^{AB}_{CD}&=&  \frac{1}{3} \left[ 
                       \delta^A_C (T^{M B}_{M D} -\frac{1}{5}\delta^B_D T^{M N}_{M N })
                      -\delta^A_D (T^{M B}_{M C} -\frac{1}{5}\delta^B_C T^{M N}_{M N }) \right. \nn\\
                     &~ & ~~~-\left.\delta^B_C (T^{M A}_{M D} -\frac{1}{5}\delta^A_D T^{M N}_{M N })
                      +  \delta^B_D  (T^{M A}_{M C} -\frac{1}{5}\delta^A_C T^{M N}_{M N }) \right]~~,
\label{sigmaabcd}
\eea
and
\beq
Y^{A B}_{C D} =  T^{A B}_{C D} - (S^{AB}_{CD} + \Sigma^{AB}_{CD}) \,\,\,.
\label{yabcd}
\eeq
%

\subsection{The Method for Catching SM Multiplets} 
\label{app-SMmultiplets}

One can apply a simple recipe to recognize how the SM multiplets are included in the large effective representations.
We will use Table \ref{multiplets}.

\begin{table} [t]
\begin{center}
\bea
SU(5)&\supset & SU(3)\times SU(2)\times U(1)\nn\\
1&=&(1,1,0)  \nn\\ 
5&=&(1,2,1/2) +(3,1,-1)  \nn\\
10&=&(1,1,1)+(\bar 3,1,-2/3)+(3,2,1/6) \nn\\ 
24&=&(1,1,0)+(1,3,0)+(3,2,-5/6)+(\bar 3, 2 ,5/6)+(8,1,0) \nn\\ 
75&=&(1,1,0)+(3,1,5/3)+(\bar 3,1,-5/3)+(3,2,-5/6)+(\bar 3,2,5/6)+\nn\\
&~& + (\bar 6,2,-5/6)+(6,2,5/6)+(8,1,0)+(8,3,0)  \nn 
\eea
\end{center}
\caption{\it The SM multiplets contained in some of the $SU(5)$ representations.}
\label{multiplets}
\end{table}

\subsubsection{SM singlets and VEVs for $\bar 5 \times 5$}
\label{app-sing-24}

The starting point is the identification of the SM multiplets in the $\bar 5$ and $5$ representations:
\beq
\bar F_A  = \underbrace{\bar 5_\alpha}_{(\bar 3, 1, -1)} + \underbrace{\bar 5_a}_{(  1, 2, 1/2)} \,\, \, , \,\, \,
F^A   = \underbrace{ 5^\alpha}_{( 3, 1, 1)} + \underbrace{5^b}_{(  1,2, -1/2)} \,\, .
\eeq

There are two SM singlets $(1,1,0)$ arising from the product $\bar F \times F$:
\beq
T^\beta_\alpha  \equiv \underbrace{ {\bar F}_\alpha }_{(\bar 3, 1, -1)} \underbrace{F^\beta}_{(3, 1, 1)} 
 =  \underbrace{ \frac{1}{3} \delta_\alpha^\beta T_\alpha^\alpha }_{(1,1,0)}  \,\,
 +  \,\,\underbrace{ T_{\alpha}^{\beta} - \frac{1}{3} \delta_\alpha^\beta T_\alpha^\alpha }_{(8,1,0)}~~~,
\eeq
and
\beq
T^b_a  \equiv \underbrace{ {\bar F}_a }_{(1, 2, -1/2)} \underbrace{F^b}_{( 1, 2,1/2)} 
 =  \underbrace{ \frac{1}{2} \delta_a^b T_a^a }_{(1,1,0)}  \,\,
 +  \,\,\underbrace{ T_a^{b} - \frac{1}{2} \delta_a^b T_a^a }_{(1,3,0)}~~~.
\eeq

We call $s^{(5)}_1$ the SM singlet related to color, and call $s^{(5)}_2$ the one related to weak isospin. 
With a normalization choice, they are
\beq
s^{(5)}_1= \frac{1}{\sqrt{3}} (T^{1}_{1}+T^{2}_{2}+T^{3}_{3}) ~~~, \,\, s^{(5)}_2= \frac{1}{\sqrt{2}} (T^{4}_{4}+T^5_5) ~~~.
\eeq
From the previous section, Eq. (\ref{eq-defs}), it turns out that 
\beq
s= \sqrt{3} s^{(5)}_1 + \sqrt{2} s^{(5)}_2 \,\, .
\eeq

Now, comparing with the previous expressions for $S_A^B$ and $\Sigma_A^B$ in Eq.~(\ref{sing-adj-ab}), and adopting a $5\times5$ matrix notation, 
it turns out that 
\beq
\langle H^{(5)}_1 \rangle = \langle S \rangle = v^{(5)}_1 \,{\rm diag}(1,1,1,1,1)\,\,\,, \,\,\, 
\langle H^{(5)}_{24}    \rangle =     \langle \Sigma \rangle = v^{(5)}_{24} \,{\rm diag}(1,1,1,-3/2,-3/2)\,\, ,
\eeq
where
\beq
v^{(5)}_1= \frac{1}{5} (\sqrt{3} s^{(5)}_1 + \sqrt{2} s^{(5)}_2) \,\,\, , \,\,\, v^{(5)}_{24}= \frac{\sqrt{2} }{\sqrt{3} \cdot 5} (\sqrt{2} s^{(5)}_1 - \sqrt{3} s^{(5)}_2)  \,\, ,
\label{eq-F-vevs}
\eeq
and we have introduced the VEV of $ H^{(5)}_1,  H^{(5)}_{24}$ to match the notation of Sec.~\ref{subsec-dynbreak}.

With these findings, one can calculate
\beq
\frac{{\rm Tr}(G_{\mu\nu} \, (\langle  H^{(5)}_{1} \rangle + \langle  H^{(5)}_{24} \rangle ) \,G^{\mu\nu} )_{G^{\rm SM}_i} }{{\rm Tr}(G_{\mu\nu}  \,G^{\mu\nu} )}= v_1^{(5)} + X_{24,i} \,v_{24}^{(5)}\,\, , \,\,\, 
\label{eq-X-5}
 \eeq
where $i=1,2,3$ corresponds to $G^{\rm SM}=\{U(1),SU(2),SU(3)\}$ and $X_{24} =\{-1/2,-3/2,1\}$. The latter results for $X_{24}$ are in agreement with Eq.~(\ref{eq-abcd}).

\subsubsection{SM singlets and VEVs for $\overline{10} \times 10$}
\label{app-sing-75}

The starting point is the identification of the SM multiplets in the $10$ and $\overline{10}$ representations. 
\beq
T^{AB}=\underbrace{10^{\alpha \beta}}_{(\bar 3, 1, -2/3)} + 
        \underbrace{10^{\alpha b}}_{(3,2,1/6)}+ 
        \underbrace{10^{a b}}_{(1,1,1)}   \,\, , \,\, \,\,\,\,
\overline{T}_{AB}=\underbrace{10_{\alpha \beta}}_{(3, 1, 2/3)} + 
        \underbrace{10_{\alpha b}}_{(\bar 3,2,-1/6)}+ 
        \underbrace{10_{a b}}_{(1,1,-1)}~~~~.
\eeq

There are three singlets $(1,1,0)$ arising from the product $T\times \overline{T}$.
The first singlet, $s^{(10)}_1$, is the one related to color only, and is obtained from
\beq
T^{\alpha \beta}_{\gamma \delta} \equiv \underbrace{T^{\alpha \beta}}_{(\bar 3, 1, -2/3)} 
\underbrace{\overline{T}_{\gamma \delta}}_{(3, 1, 2/3)}  =  
\underbrace{\frac{1}{6} (\delta^\alpha_\gamma \delta^\beta_\delta -
\delta^\alpha_\delta \delta^\beta_\gamma) 
T^{\mu \nu}_{\mu \nu} }_{(1,1,0)}  
 +  \underbrace{T^{\alpha \beta}_{\gamma \delta} -
\frac{1}{6} (\delta^\alpha_\gamma \delta^\beta_\delta -\delta^\alpha_\delta \delta^\beta_\gamma) 
T^{\mu \nu}_{\mu \nu}}_{(8,1,0)}~~~. 
\eeq
Choosing a convenient normalization, it is
\beq
s^{(10)}_1= \frac{1}{\sqrt{3}} (T^{12}_{12}+T^{13}_{13}+T^{23}_{23}) 
= \frac{1}{\sqrt{3}} T^{\mu \nu}_{\mu \nu}|_{\mu < \nu}~~~.
\eeq

The second singlet, $s^{(10)}_2$, related to both color and weak isospin, comes from the product
\beq
T^{\alpha b}_{\gamma d} \equiv \underbrace{T^{\alpha b}}_{(3, 2, 1/6)} 
\underbrace{\overline{T}_{\gamma d}}_{(\bar 3, 2, -1/6)}  =
\underbrace{
\frac{1}{6} \delta^\alpha_\gamma \delta^b_d T^{\mu n}_{\mu n}
}_{(1,1,0)} 
+ \underbrace{ 
\frac{1}{3} \delta^\alpha_\gamma T^{\mu b}_{\mu d} - \frac{1}{6} \delta^\alpha_\gamma \delta^b_d 
T^{\mu n}_{\mu n} 
}_{(1,3,0)} + ... \,\,\, ,
\eeq
and turns out to be given by
\beq
s^{(10)}_2= \frac{1}{\sqrt{6}} (T^{14}_{14}+T^{24}_{24}+T^{34}_{34}+T^{15}_{15}+T^{25}_{25}+T^{35}_{35}) 
= \frac{1}{\sqrt{6}} T^{\mu n}_{\mu n}~~~.
\eeq

Finally, the last singlet, $s^{(10)}_3$, related to weak isospin only, is obtained from 
\beq
T^{a b}_{c d} \equiv \underbrace{T^{a b}}_{(1, 1, 1)} 
\underbrace{\overline{T}_{c d}}_{(1, 1, -1)} = \underbrace{\frac{1}{2} (\delta^a_c \delta^b_d-
\delta^a_d \delta^b_c) T^{m n}_{m n}}_{(1,1,0)}~~,
\eeq
and is simply 
\beq
s^{(10)}_3=T^{45}_{45}=T^{mn}_{mn}|_{m<n}~~.
\eeq

Now, comparing with the expressions for $S, \Sigma$ and $Y$ in Eqs.~(\ref{sab}), (\ref{sigmaabcd}) and (\ref{yabcd}), and adopting a $10\times 10$ matrix notation, 
it turns out that 
\bea
\langle H^{(10)}_{1} \rangle=\langle S\rangle=v^{(10)}_1\cdot {\rm diag}(\mathbb{1}_3,\mathbb{1}_6,1) \,\,\, , \,\,\, 
\langle H^{(10)}_{24} \rangle=\langle\Sigma\rangle=6 \,v^{(10)}_{24}\cdot {\rm diag}(\mathbb{1}_3,-\frac{1}{4}\cdot \mathbb{1}_6,\frac{3}{2}) \,\,,\nn \\
\langle H^{(10)}_{75} \rangle=\langle Y\rangle=-3\,v^{(10)}_{75}\cdot {\rm diag}(\mathbb{1}_3,- \mathbb{1}_6, 3) \,\,\, , \,\,\,\,\,\,\,\,\,\,\,\,\,\,\,\,\,\,\,\,\,\,\,\,\,\,\,\,\,\,\,\,\,\,\,\,\,\,\,\,\, 
\label{eq-VEV-10}
\eea
where 
\bea
& &v^{(10)}_1 =\frac{1}{10} (\sqrt{3} s^{(10)}_1 +\sqrt{6} s^{(10)}_2 +s^{(10)}_3) \\
&&v^{(10)}_{24}=\frac{\sqrt{2}}{45\sqrt{3}} (2 \sqrt{2} s^{(10)}_1 -s^{(10)}_2 -\sqrt{6} s^{(10)}_3) \\
&&v^{(10)}_{75}=-\frac{1}{18\sqrt{3} } ( s^{(10)}_1 - \sqrt{2} s^{(10)}_2 +\sqrt{3} s^{(10)}_3)  \,\, .
\eea
and we have introduced the VEVs of $ H^{(10)}_1,  H^{(10)}_{24}$ and $H^{(10)}_{75}$ to match and clarify the notation of Sec.~\ref{subsec-dynbreak}.

Hence, one finds that, for $i=1,2,3$,
\beq
\frac{{\rm Tr}(G_{\mu\nu} \, (\langle  H^{(10)}_{1} \rangle + \langle  H^{(10)}_{24} \rangle + \langle  H^{(10)}_{75} \rangle ) \,G^{\mu\nu} )_{G^{\rm SM}_i} }{{\rm Tr}(G_{\mu\nu}  \,G^{\mu\nu} )}= v_1^{(10)} + X_{24,i} \, v_{24}^{(10)} + X_{75,i} \, v_{75}^{(10)}  \,\, , \
\label{eq-X-10}
 \eeq
where $G^{\rm SM}=\{U(1),SU(2),SU(3)\}$, $X_{24}=\{-1/2,-3/2,1\}$ and $X_{75}=\{-5,3,1\}$. These results are in agreement with Eq.~(\ref{eq-abcd}).

\subsubsection{SM singlets and VEVs for $24 \times 24$}
\label{app-sing-200}

Adopting now a $24\times 24$ matrix notation (see the appendix of Ref.~\cite{Calmet:2009hp}) the VEVs can be written as
\bea
\langle H^{(24)}_{1} \rangle= \langle S\rangle=v^{(24)}_1 \, \mathbb{1}_{24} \,\,\, , \,\,\, 
\langle H^{(24)}_{24} \rangle=\langle\Sigma\rangle= v^{(24)}_{24} \, {\rm diag}(\mathbb{1}_8,-\frac{3}{2}\cdot\mathbb{1}_3,-\frac{1}{2}, -\frac{1}{4}\cdot \mathbb{1}_{12})\,, \nn \\
\langle H^{(24)}_{75} \rangle=\langle Y\rangle = v^{(24)}_{75}\, {\rm diag}(\mathbb{1}_8, 3\cdot \mathbb{1}_3, -5,- \mathbb{1}_{12}) \,\,,\,\,\,\,\,\,\,\,\,\,\,\,\,\,\,\,\,\, \,\,\nn \\
\langle H^{(24)}_{200} \rangle=\langle Z\rangle= v^{(24)}_{200}\, {\rm diag}(\mathbb{1}_8, 2\cdot \mathbb{1}_3, 10,-2\cdot\mathbb{1}_{12})  \,\,.\,\,\,\,\,\, \,\,\,\,\,\,\,\,\,\,\,\,
\label{eq-VEV-24}
\eea

Hence, one finds that, for $i=1,2,3$,
\beq
\frac{{\rm Tr}(G_{\mu\nu} \, ( \sum_{r=1,24,75,200} \langle  H^{(24)}_{r} \rangle  ) \,G^{\mu\nu} )_{G^{\rm SM}_i} }{{\rm Tr}(G_{\mu\nu}  \,G^{\mu\nu} )}= v_1^{(24)} + X_{24,i} \, v_{24}^{(24)} + X_{75,i} \, v_{75}^{(24)} + X_{200,i} \, v_{200}^{(24)}  \,\, , 
 \eeq
where $G^{\rm SM}=\{U(1),SU(2),SU(3)\}$, $X_{24}=\{-1/2,-3/2,1\}$, $X_{75}=\{-5,3,1\}$ and $X_{200}= \{10,2,1\}$. 
These results are in agreement with Eq.~(\ref{eq-abcd}).

\subsubsection{Summary and notation}

Notice that we defined the VEVs for $r=10$ in Eq.~(\ref{eq-VEV-10}) in such a way that, for the effective representations 1 and 24, 
they precisely correspond to those of the case with $r=5$, see Eq.~(\ref{eq-X-5}). 
Similarly, we defined the VEVs for $r=24$ in Eq.~(\ref{eq-VEV-24}) in such a way that, for the effective representations 1,24 and 75, 
they precisely correspond to those of the case with $r=10$, see Eq.~(\ref{eq-X-10}). 
Hence, in order to simplify the notation, we can define
\beq
v_1 \equiv v^{(24)}_1=v^{(10)}_1=v^{(5)}_1 \,\, , \,\,\, v_{24} \equiv v^{(24)}_{24}=v^{(10)}_{24}=v^{(5)}_{24} \,\,, \,\, v_{75} \equiv v^{(24)}_{75}=v^{(10)}_{75}
\,\,, \,\, v_{200} \equiv v^{(24)}_{200}\,\, .
\label{app-VEVs}
\eeq
and summarize our results by means of the following expression
\begin{align}
&\frac{{\rm Tr}(G_{\mu\nu} \, (  \langle  H^{(R)}_{1} +H^{(R)}_{24}+ a_R H^{(R)}_{75}+ b_R H^{(R)}_{200}\rangle  ) \,G^{\mu\nu} )_{G^{\rm SM}_i} }{{\rm Tr}(G_{\mu\nu}  \,G^{\mu\nu} )}\nonumber\\
&= v_1 + X_{24,i} \, v_{24} + a_R \,X_{75,i} \, v_{75} +b_R \, X_{200,i} \, v_{200} \,\, , 
\label{eq-summ}
 \end{align}
where $a_{5}=b_{5}=b_{10}=0$, while $a_{10}=a_{24}=b_{24}=1$.

In addition, thanks to Eq.~(\ref{app-VEVs}), for any $R=5,10,24$  it is possible to write the VEVs of the effective representations $r=1,24$ using an equivalent $5\times 5$ matrix form
\beq
\langle H^{(R)}_1 \rangle \doteq v_1 \,\mathbb{1}_5 \,\, , \,\, \langle H^{(R)}_{24}\rangle \doteq v_{24}\, {\rm diag}(\mathbb{1}_3,-3/2\cdot \mathbb{1}_2) \,.
\eeq

\section{Approximate solution for GCU}
\label{App-app}

\begin{figure}[htb!]
\vskip .5cm 
 \begin{center}
 \includegraphics[width=7.6 cm]{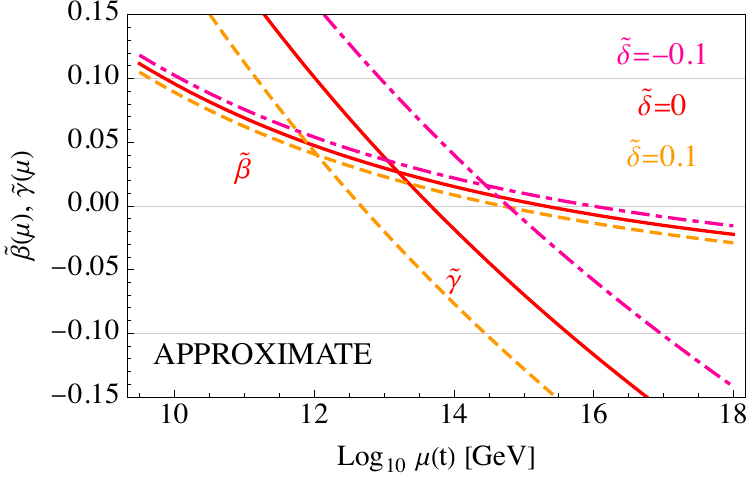}  
 \end{center}
\caption{\baselineskip=12 pt \small \it
Solid lines are $\tilde \beta(\mu)$ and $\tilde \gamma(\mu)$ according to the approximate solution, for $\tilde \delta=0, \pm 0.1$.} 
\label{fig-tbtg-app}
\vskip .5 cm
\end{figure}

As for the approximate solution, from Eq.~(\ref{eq-unif-gr-2}) and expanding to first order in $\tilde \beta$, $\tilde \gamma$ and $\tilde \delta$, 
one obtains
\bea
 f_{21}(\mu) &=& \frac{1+\alpha +\beta +\gamma +\delta } { 1+ \alpha + 3 \beta -\frac{3}{5}\gamma  +\frac{1}{5}\delta }  
 \approx  1 - 2 {\tilde \beta} + \frac{8}{5} \tilde \gamma   + \frac{4}{5} \tilde \delta \,\,\, \nonumber \\
 & & \\
 f_{31}(\mu)  &=& \frac{1+\alpha +\beta +\gamma +\delta } {1+ \alpha -2 \beta -\frac{1}{5}\gamma  +\frac{1}{10}\delta}  
 \approx 1 +3 {\tilde \beta}+ \frac{6}{5} \tilde \gamma +  \frac{9}{10} \tilde \delta \nonumber \,.
\label{eq-sys}
\eea
The system of two equations above can now be solved for any value of $\mu=M_X$, with redundancy of solutions, as there are three unknowns
 ($\tilde \beta$, $\tilde \gamma$ and $\tilde \delta$), 
\beq
\tilde \beta = -\frac{2}{3} \left(   \frac{f_{21}-1}{4} - \frac{f_{31}-1}{3}  +\frac{1}{10} \tilde \delta \right) \,\,\, , \,\,\, 
\tilde \gamma = \frac{5}{36}  \left( 3 f_{21} + 2 f_{31} -5 -\frac{21}{5} \tilde \delta \right) \,.
\label{eq-tbtg-app}
\eeq
For fixed values of $\tilde \delta$, we compare the exact and approximate solutions displayed in Fig.~\ref{fig-tbtg} and Fig.~\ref{fig-tbtg-app} respectively: they agree pretty well. 
The comparison can be used to check the consistency of the perturbative approach.

\section{A (toy) UV completion for non-renormalizable operators}
\label{apptoy}

In this paper we have used the dynamical breaking of $SU(5)$ to impose conditions on GCU via the $d=5$ operators given by Eq.~(\ref{eq-opgr}). 
To this end we are using the higher dimensional operator in Eq.~(\ref{eq:DTS}) from the $\langle \bar F_R F_R \rangle$ condensate (the DTS operator), along with the non-renormalizable operators (\ref{eq-F}), (\ref{eq-T}) and (\ref{eq-A}) from the respective condensates $\langle \bar F F\rangle$, $\langle \bar T T\rangle$ and $\langle \bar A A\rangle$ (kinetic operators). In this section we provide some toy UV completions that can give rise to such operators at the condensation scale.

\subsection{The DTS operator}
\label{app:DTS}

Here we are considering the $d=5$ operator
\beq
\mathcal L_{\rm eff}=-\frac{c_R}{\Lambda} H_5^\dagger ( F_R \bar F_R) H_{5}  \, ,
\label{eq:effHFFH}
\eeq%
where $\Lambda$ is the cutoff of the theory, giving rise to the DTS.
In our theory we assume that the $SU(5)$ GUT symmetry is dynamically broken when a condensate is formed at some scale $\Lambda_G$, when the confining group $G$ is strongly coupled. 

Let us start by considering the case $R=5$.
The relevant fields are the fermions, $F(5,R_G)$, $\bar F(\bar 5,\bar R_G)$ (notice that we need both $F$ and $\bar F$ to cancel anomalies), 
the Higgs field $H_5(5,1)$, and the $SU(5)$ gauge bosons. 
As the breaking is dynamical we do not have any scale in the theory, while the only scale $\Lambda_G$ is dynamically created by the strong force in $G$. In order to generate a mass term for the Higgses we introduce heavy fermions, transforming as a representation $R_G$ of $G$ and singlets under $SU(5)$, $\psi(1,R_G)$ and $\bar \psi(1,\bar R_G)$, with a mass $M$, a bit larger than $\Lambda_G$ to not change the $G$ dynamics at the condensation scale, so that the field $\psi$ can be integrated out. The renormalizable Yukawa Lagrangian is then
\beq
\mathcal L=h_F \, \bar F_A  H_5^A \psi +h_F^* \, \bar \psi  (H^\dagger_5)_A F^A +M \bar \psi \psi \, ,
\eeq
where $h_F$ are the Yukawa couplings and $A=1,...,5$ refers to $SU(5)$.
\newline\indent
After integrating out $\psi$ the Lagrangian (at scales $\mu<M$) is written as
\beq
\mathcal L_{\rm eff}=-\frac{|h_F|^2}{M} H_5^A ( \bar F F)_A^B\, (H^\dagger_5)_B \, , 
\eeq
which reduces to (\ref{eq:effHFFH}) by the identification $c_F=|h_F|^2$ and $\Lambda \sim M$. 
For scales below the condensation scale the effective Lagrangian can be written as
\beq
\mathcal L_{\rm eff}=-\lambda_F^2\,\Lambda_G \,H_5^\dagger \left(\langle H_{1}^{(5)}+H_{24}^{(5)}\rangle\right)H_5, \quad \textrm{with}\quad 
\lambda_F^2\equiv |h_F|^2 \Lambda_G/\Lambda\,.
\eeq
Notice that this mechanism is similar to that giving rise to the Weinberg operator in the leptonic sector, after integrating out the heavy right-handed neutrinos. After condensation $\langle \bar F_R F_R \rangle$, the previous Lagrangian gives a mass to the Higgs five-plet. 

In the case $R=10$, the condensation fermions are $T(10,R_G)$ and $\bar T(\overline{10},\bar R_G)$. 
The argument is similar, with the only difference being that the heavy fermions $\psi$ and $\bar\psi$ cannot be singlets under $SU(5)$, but should rather transform as $\psi(5,R_G)$ and $\bar \psi(\bar 5,\bar R_G)$. The Lagrangian becomes 
\beq
\mathcal L=h_{T} \, \bar T_{AB}  H_5^A \psi^B +h_{T}^* \, \bar \psi_A  (H^\dagger_5)_B T^{AB} +M \bar \psi_A \psi^A \, ,
\eeq
giving, upon condensation of $\langle \bar T\, T \rangle=H_{1}^{(10)}+H_{24}^{(10)} +  H_{75}^{(10)} $, 
\beq
\mathcal L_{\rm eff}
=-\frac{|h_T|^2}{M} H_5^A \langle      H_{1}^{(10)}+H_{24}^{(10)}  \rangle_A^B\, (H^\dagger_5)_B  \, .
\eeq
As $\bar 5\times5=1+24$, the term with the $75$ provides indeed no contribution.

Finally, in the case $R=24$, where  the condensation fermions are $A(24,R_G)$ and $\bar A(24,\bar R_G)$. the above arguments are reproduced by taking for instance $\psi(\bar 5,R_G)$ and $\bar \psi(5,\bar R_G)$.

\begin{figure}[htb!]
\vskip .5cm 
 \begin{center}
 \includegraphics[width=14 cm]{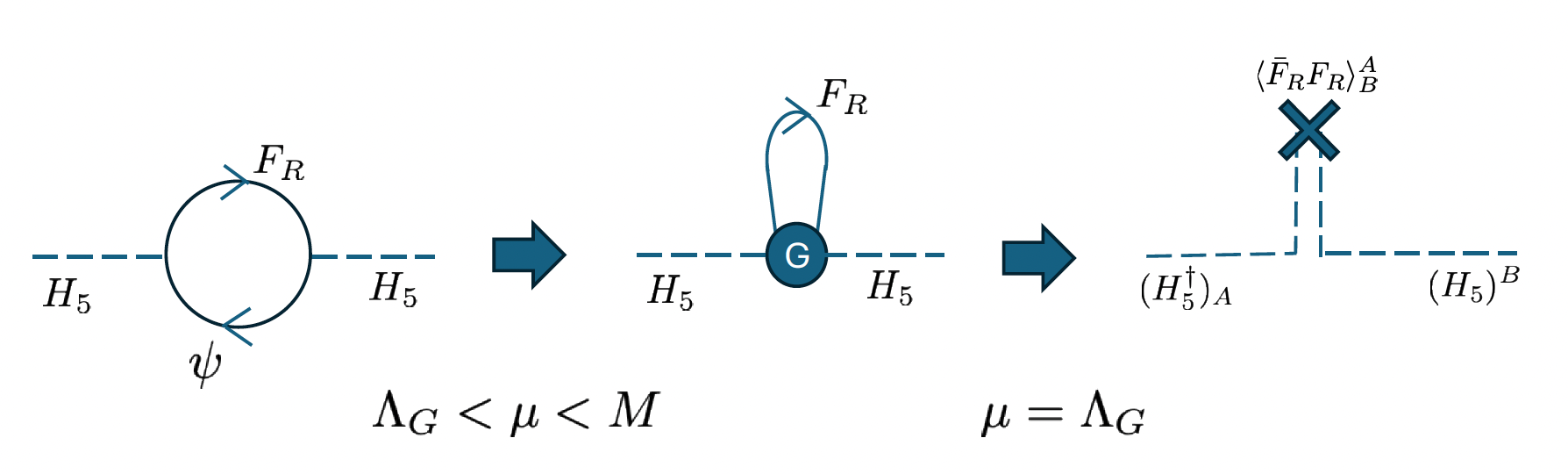}
 \end{center}
\caption{\baselineskip=12 pt \small \it
Toy model's graphical representation of the condensation and the generation of the non-renormalizable operator providing the DTS.} 
\label{fig-graph-dts}
\vskip .5 cm
\end{figure}

Fig.~\ref{fig-graph-dts} provides a graphical representation for the condensation mechanism within the toy model.

\subsection{The higher dimensional kinetic operators}
\label{app:kinetic}

The modification of GCU in models with $SU(5)$ dynamical breaking is induced via the $d=7$ operators
\beq
 -\frac{1}{4} \frac{\bar c_R}{\Lambda^3} {\rm Tr}(G_{\mu\nu} \,  \bar F_{R}  F_{R}  \,G^{\mu\nu} ) \,.
  \label{eq:GCUapp}
\eeq
We will now see how can we generate such term in the Lagrangian.

We introduce heavy fermions, singlets under $G$ (so they cannot condense at the scale $\Lambda_G$), and transforming under the representation $R$ of $SU(5)$, $f(R,1)$ and $\bar f(\bar R,1)$, with mass $M_f$, and a heavy real scalar $\varphi$ with a mass $M_s$ and zero vacuum expectation value, with a Yukawa coupling and Lagrangian
\beq
\mathcal L=\varphi \bar ff+M_f \bar ff-\frac{1}{2}M^2_s \varphi^2\quad \Rightarrow\quad m_f(\varphi)=M_f+\varphi \,.
\eeq 
We are assuming the mass $M_s$ to be larger than $\Lambda_G$ for the scalar field $\varphi$ to be integrated out before the condensation dynamics.

The fermion $f$ contributes to the renormalization of the $SU(5)$ kinetic term. At one-loop the kinetic Lagrangian is~\cite{Shifman:1979eb}
\beq
\mathcal L_{kin}=-\frac{1}{4}G_{\mu\nu}G^{\mu\nu}\frac{b_f g_5^2}{16\pi^2}\log\frac{\Lambda^2}{m_f^2(\varphi)} \,,
\eeq
where $b_f$ is the coefficient of the beta-function contribution from the fermion $f$ and $g_5$ the $SU(5)$ gauge coupling. 
The previous Lagrangian can be expanded in a power series in $\varphi$ and yields
\beq
\mathcal L_{\rm eff}=\frac{1}{4}\,\frac{\alpha_5 \, b_f}{2\pi}\, \frac{\varphi}{M_f}G_{\mu\nu}G^{\mu\nu} \,.
\eeq

We now introduce the Yukawa coupling of $\varphi$ with the confining fermions as
\beq
\Delta \mathcal L=\lambda_R \varphi \bar F_R F_R \,,
\eeq
and integrate out the scalar $\varphi$ as $\varphi=-\lambda_R(\bar F_R F_R+\bar ff)/M_s^2$, which yields
\beq
\mathcal L_{\rm eff}=-\frac{1}{4}\, \frac{\alpha_5 \lambda_R b_f}{2\pi} \frac{\bar F_R F_R}{M_f M_s^2}\, G_{\mu\nu} G^{\mu\nu} \, ,
\eeq 
which can then be identified with (\ref{eq:GCUapp}) assuming $M_f\sim M_s$, $M_f M_s^2\equiv \Lambda^3$ and 
\beq
\bar c_R \equiv \frac{\alpha_5 \lambda_R b_f}{2\pi}\,  \,.
\eeq
Finally, at the condensation scale $\mu\sim \Lambda_G$ the effective Lagrangian can be written as in Eq.~(\ref{eq:kinetic}) with 
\beq
c_R \equiv \bar c_R\, \left(\Lambda_G/\Lambda \right)^3 \,.
\eeq

\begin{figure}[htb!]
\vskip .5cm 
 \begin{center}
 \includegraphics[width=14 cm]{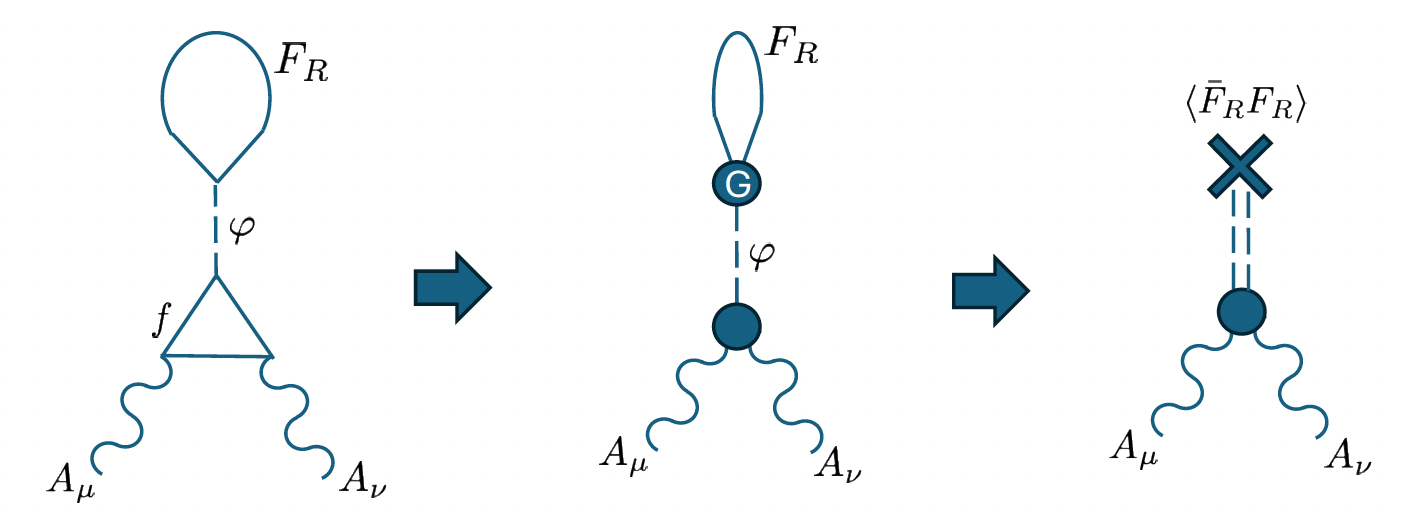}
 \end{center}
\caption{\baselineskip=12 pt \small \it
Toy model's graphical representation of the condensation and the generation of the non-renormalizable kinetic operator in (\ref{eq:GCUapp}).} 
\label{fig-graph-dtss}
\vskip .5 cm
\end{figure}

Notice that this effective Lagrangian is similar to that giving rise to the decay $h\to \gamma\gamma$ in the SM, where the scalar field $\varphi$ plays the role of the Higgs $h$, and the heavy fermion $f$ that of the top-quark $t$. Fig.~\ref{fig-graph-dtss} provides a graphical representation for this mechanism within the toy model.

\bibliographystyle{elsarticle-num} 
\bibliography{bib-unif} 
\end{document}